\documentclass[12pt]{article}
\usepackage{amsfonts}
\usepackage{graphicx}
\newcommand{\beqal}[1]{\begin{eqnarray}
\label{#1}}
\newcommand{\eeqa}{\end{eqnarray}}
\newcommand{\ket}[1]{\vert#1\rangle} 
\newcommand{\braket}[2]{\langle#1\vert#2\rangle} 
\newcommand{\obraket}[3]{\langle #1 \mid #2 \mid #3 \rangle}
\begin{document}

\title{\Large Non-spherical multicluster approximation in light nuclei}
\author{A. Gij\'on, F.J. G\'alvez, F. Arias de Saavedra and  E. Buend\'{\i}a \\
Departamento de F\'{\i}sica At\'omica, Molecular y Nuclear, \\
Facultad de Ciencias, Universidad de Granada, \\E-18071 Granada, Spain}
\date{ }
\maketitle

\noindent

\begin{abstract}
Multicluster models consider that the nucleons can be moving around
different centers in the nuclei. These models have been widely used to
describe light nuclei but always considering that the mean field is
composed of isotropic harmonic oscillators with different centers. In
this work, we propose an extension of these models by using
anisotropic harmonic oscillators. The strenghts of these oscillators, 
the distance among the different centers and the disposition of the
nucleons inside every cluster are free parameters which have
been fixed using the variational criterion. We have used a hamiltonian
with the kinetic energy terms and a phenomenological two-body
potential like Volkov V2 potential. All the one-body and two-body
matrix elements have been analytically calculated. Only a numerical
integration on the Euler angles, it is needed to carry out the projection on the values of
the total spin of the state and its third component. We have studied the ground
state and the first excited states of $^8$Be, $^{12}$C and $^{10}$Be
getting good results for the energies. The disposition of the nucleons
in the different clusters have been also analyzed by using projection on
the different cartesian planes getting much more information than when
the radial one-body density is used. 
\end{abstract}


\section{Introduction.} 

In light nuclei, the differences of the bound energy among deuteron,
triton and, mainly, the $\alpha$ particle,  are arguments that allow to
consider that nucleons in the stationary states can be located around
more than one point in the space forming clusters of nucleons, being
the $\alpha$ particle the main way of clustering. Microscopic
multicluster models have widely developed this form of describing the
structure of stationary states since Wheeler proposed the Resonating
Group Model (RGM) up to describe the ground state of $^8$Be as a pair
of separated $\alpha$ particles moving around each other \cite{Whee-1937}.  
Four years after, Margenau \cite{Marg-1941} presented an alternative to
RGM that avoided some of its antisymmetrization problems by describing
the eigenstates using Slater determinants with single particle
wave functions centered around several fixed points in space.  This
proposal is generalized by Griffin and Wheeler with the Generator
Coordinate Method (GCM) that leaves free the positions of the centers
of the clusters \cite{GrWh-1957}. The previously cited works joined to the
further systematization of GCM by Brink \cite{Brin-1967,BrWe-1968} are the
background of all the models that have utilized the idea of different
clusters of nucleons within nuclei. The most general models eliminate
the restriction that clusters are only formed by $\alpha$ particles, and
drive to a molecular vision of nuclei \cite{AHT-1973} and are known as
$\alpha$-cluster or multicluster models. 

Multicluster models in their different versions have been widely used
to study different aspects of the dynamics of light and medium
nuclei. Important results on the structure and spectroscopy of the
stationary states of nuclei up to $^{40}$Ca have been achieved
\cite{KHO-1995,DeDu-1997,DuDe-1997}, moreover they are of great
utility to describe nuclei rich in neutrons specially when
Antisymmetric Molecular Dynamics (AMD) is used 
\cite{KaHo-1995,KHD-1999,KaHo-2002}. The structure of multicluster
vectors have also allowed to obtain cuantitative results in the
description of nuclear reactions that involve light and medium nuclei 
\cite{LKT-1981,Desc-2004,DuDe-2008}.
A complete presentation of the theoretical aspects of this kind of
approximations and of the specific applications can be found in the
work by Descouvemot and Dufour \cite{DeDu-2012} or in the one by von 
Oertzen, Freer and Kanada-En’yo \cite{OFK-2006}.

Multicluster models can be considered as an independent particle approximation in which
nucleons move in a mean potential with more than one minimum in
contrast with the usual model where nucleons move in a local or
non-local one-particle potential with a well defined minimum and with
spherical symmetry. This last one corresponds to the simplest
multicluster model. In almost all the multicluster models used up to
now, harmonic oscillators potentials spherically symmetric respect to
their centers are used except in the case of a sigle center  where
deformed harmonic oscillator potentials have also been used.  The
distances among the centers of the clusters and the oscillator
parameters are variationally fixed in the simplest approximations or all
the posible values of the relative distances among the clusters are 
mixed in methods based in GCM
\cite{DuDe-1996}. The basic element in these approximations is a
determinant or a linear combination of determinants built by using the
single particle states around the different centers that describe the
intrinsic form of the nuclei. This corresponds to a generating
state. The vectors used to approximate the physical state are obtained
from the generating ones after a projection on the subspace with
total angular momentum, parity and total isospin values corresponding to the
state to be described. This is the usual scheme of this kind of
approximations and is not conditioned by the symmetry of every of the
potentials that confine each of the clusters.

Within the multicluster approximation, the relative position of the
center of the different clusters has influence on the symmetry of the
potential felt by the nucleons in every cluster. This property
does not make very adequate the use of spherically symmetric potentials respect
to its center. Nevertheless, the possibility of working with non-spherical
monoparticular states has not been explored for this kind of
models even tough this does not involve an excesive difficulty from
the technical point of view. The main goal of the present work is to
analyze the multicluster approximation when non-spherical or deformed harmonic
oscillator potentials centered in different points of the space are
used as mean field. Within our anisotropic multicluster model, there is
also no restriction neither in the number of clusters used nor in the
number of nucleons inside every cluster. The calculation of the matrix
elements in their spatial part can be analytically performed and only
a numerical approximation is needed for proyecting to defined
total angular momentum  since projection on total isospin and parity can be
algebraically carried out. We want to test the possibilities of the
proposed method by studying the low energy states of $^8$Be, 
$^{10}$Be and $^{12}$C, exploring different
dispositions of the nucleons in the clusters.

\section{Anisotropic multicluster model (AMM).}

The AMM is an independent particle model
approximation built on a mean field constructed from a set of anisotropic
harmonic oscillators centered at different spatial points. The
distribution of nucleons around each center, the distances among the
centers and the values of the harmonic oscillator parameters are degrees
of freedom of our model that will be fixed by using the variational
criterion. The eigenstates of this potential with several centers will not be
determined, instead they will be approximated by the product of 
the eigenstates of every of the harmonic
oscillators using them as they were independent. So, around to every
one of the centers we use as monoparticular vectors the set
\begin{equation}
\label{e1.1}
\braket{\vec{q}}{\psi_{\mu}} = 
\prod_{k=1}^3 
\left(\frac{\alpha_{\mu,k}^{2n_{\mu,k}+1}}{\Gamma(n_{\mu,k}+ \frac 1 2)} \right)^{1/2}
(x_k-a_{\mu,k})^{n_{\mu,k}}
e^{-\frac 1 2 \alpha_{\mu,k}^2 (x_k-a_{\mu,k})^2}
\ket{m_{s,\mu}}\ket{m_{t,\mu}} \, \, ,
\end{equation}
where $\mu$ indicates all the elements that define the
single particle vector: the coordinantes of its center, $a_{\mu,k}$,
the harmonic oscillator parameters, $\alpha_{\mu,k}$, the monomia in
the vector, $n_{\mu,k}$, and the third components of spin, $m_{s,\mu}$
and isospin, $m_{t,\mu}$. These vectors span the same subspace than the
corresponding eigenstates of the anisotropic harmonic oscillator. 
Since this set is more operative, it will  be used instead of the
corresponding orthogonal basis. 

Once the nuclei and the state to be described are fixed, and the number
of clusters and the distribution of the nucleons in the different
clusters are chosen, we can use
these single particle vectors to build one or more Slater determinants to
describe the intrinsec form of the nuclei. The number of determinants
that can be built corresponds
to the possibility of permuting the orientations of the third
component of isospin of
the different monoparticular states. This does not affect to the
structure of the cluster, since isospin degrees of freedom are
decoupled from the rest, spatial and spin, of them.  
So using  the resulting determinants, it  is
possible to build states with defined total isospin and all of them
with the same spatial and spin structures.

Using these intrinsic vectors, linear combinations of them are used to
approximate the considered nuclear state with defined parity and total
angular momentum since total isospin have been already taken into
account. The obtention of states with well defined parity is trivial, 
once the center of mass is established,  just considering the effects of parity operator on the
centers and single particle vectors. The last step is to get well
defined total angular momentum, this can be obtained by applying the 
Peierls--Yoccoz projector for the group of rotations. So the state
vector to approximate the nuclear state under study can be written as
\begin{equation}
\label{e1.3}
\ket{\Psi^{\pi=\pm}_{KJM,TM_T}}  =  P^J_{KM}\frac{1}{\sqrt{2}}
\Bigl(\ket{\Psi_{TM_T}}\pm P\ket{\Psi_{TM_T}}\Bigr) =
P^J_{KM} \ket{\Psi_{TM_T}^\pi} \, \, .
\end{equation}
$\ket{\Psi_{TM_T}}$ represents the linear combination of generating
determinants with different isospin orientations need to get states
with  total isospin, $T$, and third component, $M_T$. 
This linear combination often reduces to only one
determinant. $P$ is the parity operator and $\pi=\pm$ is the parity of
the state. $P^J_{KM}$ is the Peierls--Yoccoz projector on states with
total angular momentum, $J$, third component $M$ in the laboratory
system and third component $K$ in the intrinsic system of the
nucleus. This can be written as
\begin{equation}
\label{e1.4}
P^J_{KM}  =  \frac{2J+1}{8\pi^2}\int_0^{2\pi}d\theta_1\int_0^{2\pi}d\theta_3
\int_0^{\pi}d\theta_2 \sin(\theta_2)\mathcal{D}^{J}_{K,M}(\theta_1,\theta_2,\theta_3)^*
R(\theta_1,\theta_2,\theta_3) \, \, , 
\end{equation}
where $\theta_1,\theta_2,\theta_3$ are the  Euler angles,
$\mathcal{D}^{J}_{K,M}(\theta_1,\theta_2,\theta_3)$ denotes the rotation
matrix and $R(\theta_1,\theta_2,\theta_3)$ the corresponding rotation
operator. That is, the projection is obtained by rotating the
generating state and integrating on all the angles weighted with the
rotation matrix. It should be noted that the rotations act on both
the spatial and spin degrees of freedom. In the projection, the
quantum number $K$ labels a set of rotational states that form a
rotational band \cite{EiGr-1988}. The allowed values of $J$ and $K$ depend on the
spatial distribution of the centers of the clusters, that is, they
depend on the symmetric group that contains all the symmetries of the
configuration considered. 

\section{Nucleon--nucleon interaction.}

The model is completed establishing the interaction between pairs of
nucleons. The characteristics of the proposed model obliges us to use
phenomelogical interactions up to obtain reasonable results. The
parameters in the phenomenological interactions should be fitted to
reproduce some of the experimental bound energies as well as the root mean square 
radii of the ground state of a set of nuclei. Some examples of this
kind of interactions are the ones proposed by Volkov \cite{Volk-1965},
Brink and Boecker\cite{BrBo-1967} and the one known as 
Minnesota interaction \cite{TLT-1977}. In order to be able to
calculate the matrix elements analytically, it is convenient that the
radial dependence of the different channels in the interaction are
parametered in terms of gaussians. In this work, we have chosen the
Volkov V2 interaction including a spin--orbit term, as in the Minnesota
interaction, that has been fixed to get a good estimation of the
excitation energy of the first $\frac {3}{2}^-$ state of
$^{15}$N \cite{DeDu-2012}, although there is no contribution from this
term in the energy for the nuclei studied in this work. 
The Coulomb interaction between pairs of
protons has been also included. So the interaction between
pairs of nucleon can be written as 
\begin{equation}
\label{e1.5}
V(i,j)  =  V^{N,v4}(i,j)+V^{N,so}(i,j)+V^{C}(i,j) \, \, ,
\end{equation}
where
\begin{equation}
\label{e1.5a}
V^{N,v4}(i,j)  =  \sum_{k=1}^2 V_k e^{-(r_{ij}/a_k)^2}
((1-M)-MP_{i,j}^{\sigma}P_{i,j}^{\tau}) \, \, ,
\end{equation} 
and $V_1=-60.65~{\rm MeV}$, $V_2=61.14~{\rm MeV}$, $a_1=1.80~{\rm
  fm}$, $a_2=1.01~{\rm fm}$, 
$M=0.6$ and $P_{i,j}^{\sigma}$ and $P_{i,j}^{\tau}$ are the spin
and isospin exchange operators, respectively. We can see that this
potential has only Wigner and Majorana parts different from zero. 
The spin--orbit interaction is equal to 
\begin{equation}
\label{e1.5b}
V^{N,so}(i,j)  =  \frac{S_0}{\hbar^2r_0^5}(\vec{r}_{ij}\times\vec{p}_{ij})
\cdot(\vec{s}_i+\vec{s}_j) e^{-(r/r_0)^2} \, \, ,
\end{equation} 
where $S_0=30~{\rm MeV~fm^5}$  and $r_0=0.1~{\rm fm}$. The Coulomb
interaction between puntual protons is 
\begin{equation}
\label{e1.5c}
V^{C}(i,j)  = 
\frac{e^2}{r_{ij}}(\frac 1 2 +\tau_{i,z})(\frac 1 2 +\tau_{j,z}) \, \, ,
\end{equation}
being $\tau_{k,z}$ the third component of the isospin operator. Since
the radial dependence of this interaction is not gaussian, we shall
use the integral transform
\begin{equation}
\label{e1.5c1}
\frac{1}{r_{ij}}  =  \frac{1}{\sqrt{\pi}}
\int_{-\infty}^{\infty} dt e^{-t^2r_{ij}^2} \, \, .
\end{equation}
The practical calculation of this part of the interaction requires to make 
use a discrete set of values of $t$ to approximate this last integral.

\section{Expectation value of operators.}

The calculation of the expectation value of one-- and two--body
operators between the proposed state vectors is the main technical
difficulty that appears when the multicluster model is used. Even in
 the presence of the angular momentum proyectors, it is still possible to
get expressions similar to the ones obtained in independent particle
models when single particle states are not orthogonal. After that the
projection on the angular degrees should be carried out.

Let us first note that for every operator in the spatial and spin
spaces that can be expressed as a spherical tensor or rank $j$, $Q_{jm}$,
we can write 
\begin{eqnarray}
\obraket{\Phi^{\pi^\prime}_{K^\prime J^\prime M^\prime ,T^\prime M^\prime_T}}{Q_{jm}}
{\Psi^{\pi}_{KJM,TM_T}} & = & \frac{(2J+1)}{8\pi^2}
\delta_{T,T^\prime}\delta_{M_T,M^\prime_T}\delta_{\pi,\pi^\prime}\nonumber\\
& & \sum_{m^\prime}\braket{JMjm}{J^\prime M^\prime}
\braket{J(K^\prime-m^\prime)jm^\prime}{J^\prime K^\prime}\nonumber\\
& & \hspace*{-1.4cm} \int d\Omega {\cal D}^{(J)*}_{K^\prime-m^\prime,K}(\Omega)
\obraket{\Phi_{T^\prime M_T^\prime}^{\pi^\prime}}{Q_{jm^\prime}R(\Omega)}{\Psi_{TM_T}^\pi}  \, .
\label{e1.6}
\end{eqnarray}
This expression simplifies for zero rank tensors such as the hamiltonian
and the mean squared radius. The integral on the non angular degrees
of freedom can be written as a sum of the expectation value of the
operator between pairs of different Slater determinants with
single particle vector not orthogonal each other. 
Let us denote by $\ket{\Phi_{T^\prime M_T^\prime}^{\pi^\prime}}$ and
$\ket{\Psi_{TM_T}^\pi}$, two A-particle Slater determinants built with the
single particle vectors $\left\{\ket{\phi_{\alpha}}\right\}$ and
$\left\{\ket{\psi_{\beta}}\right\}$, respectively. The matrix elements
of fully simmetric under the exchange of particles one--body and
two--body operators, $O_1=\sum_{i}O_1(i)$ and  $O_2=\sum_{i\neq
  j}O_2(i,j)$ respectively, can be written as
\begin{eqnarray}
\label{e1.7a}
\obraket{\Phi}{O_1(1,\ldots,A)}{\Psi} & = & |B|\sum_{\alpha,\beta}
\obraket{\phi_{\alpha}}{O_1(1)}{\psi_{\beta}} B^{-1}_{\beta,\alpha} \,
\, , \\
\obraket{\Phi}{O_2(1,\ldots,A)}{\Psi} & = & |B|
\sum_{\alpha_1,\alpha_2;\beta_1,\beta_2}
\obraket{\phi_{\alpha_1}\phi_{\alpha_2}}{O_2(1,2)}
{\psi_{\beta_1}{\psi_{\beta_2}}}\nonumber\\
& & \left(
B^{-1}_{\beta_1,\alpha_1}B^{-1}_{\beta_2,\alpha_2}
-B^{-1}_{\beta_1,\alpha_2}B^{-1}_{\beta_2,\alpha_1}\right) \, \, ,
\label{e1.7b}
\end{eqnarray}
where $|B|$ is the determinant of the overlap matrix,
$B=(\braket{\phi_{\alpha}}{\psi_{\beta}})$, and
$B^{-1}_{\alpha,\beta}$ is the $\alpha,\beta$ element of the inverse
matrix. The matrix elements between the single particle vectors are the
basic quantities to be determined, they are three-- and six--dimension
spatial integrals that can be analytically solved in all the cases
involved in our multicluster method. The way of solving these integral
is discussed in the Appendix and is based on a recurrence relation.

\section{Results.}

We have chosen the nuclei $^8$Be, $^{10}$Be and $^{12}$C to illustrate
the importance of leaving free the three parameters in the harmonic
oscillator for every of the centers considered. $^8$Be and $^{12}$C
are the two light nuclei more studied with different methods including
the multicluster one while $^{10}$Be is an example of rich neutron
nucleus that allows different geometries in the clusters and different
ways of disposing the nucleons inside the clusters. 

In the cases of $^8$Be and $^{12}$C, we will study the two more
obvious distributions for every nucleus, that is, to consider that
either all the nucleons move
around a single center or that the nucleons are disposed forming
$\alpha$-clusters, two for berilium and three for carbon. In this last
case, we will study the cases when the three $\alpha$-clusters form an
equilateral triangle, an isosceles one and when they are aligned. 

Hereafter, when we say spherical or isotropic approximation, we will refer to the
case when the three
parameters in the harmonic oscillator are equal for every of the
clusters considered to build the single particle states.  On the
contrary, we will say deformed or anisotropic approximation when the three 
oscillator strenghts can have different values for every
cluster. Moreover, we use a Gauss--Hermite integration rule with 30
points for every angle in the numerical integration on the three Euler
angles since this choice garantees no loss of precision in the digits
of the results to be shown. 

\begin{table}
{\scriptsize
\begin{center}
\begin{tabular}{|l|llcr|llcr|r|}
\hline
$J^{\pi},K$ & $(\alpha_x,d_y,d_z)$ & $\Delta$ & $\sqrt{<r^2>}$ &  $E$ &
 $(\alpha_x,d_y,d_z)$ & $\Delta$ & $\sqrt{<r^2>}$ &   $E$ & $E_{exp}$\\
\hline
 & \multicolumn{7}{c}{1 center} & & \\
\hline
$0^+,0$ &  0.62,1.,1.   & - & 2.16 & -39.20 &  
0.76,1.,0.63 & - & 2.32 & -50.46 & -\\ 
$2^+,0$ &               & - & 2.16 &   1.95 &
             & - & 2.34 &   3.26 & 3.04 \\
$4^+,0$ &               & - & 2.16 &   6.49 &
             & - & 2.38 & 11.67 & 11.40\\
\hline
 & \multicolumn{7}{c}{2 centers} & & \\
\hline
$0^+,0$ & 0.73,1.,1.  & 3.67 & 2.41 & -53.20 & 
0.76,1.0.87 & 3.62 & 2.41 & -54.04 & -\\
$2^+,0$ &             &      & 2.42 &   3.37 &
            &      & 2.42 &   3.49 & 3.04\\
$4^+,0$ &             &      & 2.45 &  11.98 &
            &      & 2.46 &  12.24 & 11.40\\
\hline
\end{tabular}
\caption{\label{tbl1}{\scriptsize $^{8}Be$ nucleus. All the results
    shown corresponds to states with $T=0$. We use one or two clusters
    (centers). On the left, an isotropic harmonic oscillator
    potential is used while on the right, a defomed one. Excitation
    energies are refered to the corresponding ground state energy and
    compared with the experimental results. All energies
    are in MeV, the oscillator parameter, $\alpha_x$, in fm$^{-1}$,
    the distance between cluster centers,   $\Delta$, and the root
    mean squared radii in fm.}}
\end{center}
}
\end{table}

In Table 1, we show the results obtained for $^8$Be. In the one-center
approximation, we have used the configuration $[(0,0,0)^4,(0,0,1)^4]$
that provides a intrinsic form that is deformed along the $z$--axis
 even for isotropic harmonic oscillator. These parameters 
are presented in all Tables using $\alpha_x$
and $d_i=\alpha_i/\alpha_x$ with $i=y,z$, that is, the strenght on
the $x$-azis and the relative deformations in the other axes respect to
the $x$-one. In the one-center case, we
compare the results without deformation in the potential (on the left)
with those from the deformed potential (on the right). Leaving free the
oscillator parameters reforces the axial deformation along the
$z$-axis with a less confinant oscillator in this axis. The increase
in the bound energy of the ground state, state with $J^\pi=0^+$, is
very important, of the order of $25 \%$ of the total energy. Moreover,
a better agreement is obtained for the excitation energies of the
states  $J^\pi=2^+,4^+$ that are associated to the rotational band of
the ground state. It is also remarkable the change in the value of the
root mean squared radius that increases when the deformed oscillator is
considered. In the case of considering the nucleons forming two
$\alpha$ particles separated a distance, $\Delta$, in the $z$-axis, we find that the
increase of the ground state bound energy caused by the deformation of
the harmonic oscillator potential is quite small, only $0.8$ MeV. The change in
the excitation energy of the studied states is also  few relevant so
both approximations in this case are almost equivalent. There are also
very small changes in the root mean squared radius and in the distance
between the clusters. The equilibrium values of the oscillator
parameters have a more important change but smaller if we compare it
to the modifications in the one center case.

\begin{table}
{\scriptsize
\begin{center}
\begin{tabular}{|l|llcr|llcr|r|}
\hline
$J^{\pi},K$ & $(\alpha_x,d_y,d_z)$ & $\Delta$ & $\sqrt{<r^2>}$ &  $E$ &
 $(\alpha_x,d_y,d_z)$ & $\Delta$ & $\sqrt{<r^2>}$ & $E$ & $E_{exp}$\\
\hline
 & \multicolumn{7}{c}{1 center} & & \\
\hline
$0^+,0$ &  0.64,1.,1.   & - & 2.23 & -76.06 &  
0.63,0.85,1.27 & - & 2.29 & -84.06 & -\\ 
$2^+,0$ &               & - & 2.23 &   1.87 &
               & - & 2.30 &   3.09 &  4.44 \\
$4^+,0$ &               & - & 2.23 &   6.21 &
               & - & 2.32 &  12.33 & 14.08\\
$0^+,0:l$ & 0.58,1.,1. & - & 2.64 & 27.01 &  
0.73,1.,0.59 & - & 3.08 & 15.58 &  7.65\\ 
\hline 
 & \multicolumn{7}{c}{3 centers} & & \\
\hline 
 & \multicolumn{7}{c}{equilateral triangle} & & \\
\hline 
$0^+,0$ & 0.73,1.,1.  & 2.63 & 2.30 & -86.56 & 
0.73,0.84,1.11 & 2.47 & 2.33 & -89.17 & -\\
$2^+,0$ &             &      & 2.31 &   2.55 &
            &      & 2.33 &  3.39 &  4.44\\
$4^+,0$ &             &      & 2.32 &  10.15 &
            &      & 2.35 & 13.19 & 14.08\\
$3^-,3$ &  0.72,1.,1. & 3.12 & 2.48 &  10.18 & 
0.66,1.07,1.18 & 3.06 & 2.49 &  11.40 & 9.64\\
\hline
 & \multicolumn{7}{c}{isosceles triangle} & &\\
\hline
$0^+,0$ & 0.76,1.,1. & 2.30 &      &        & 
0.76,0.87,1.08 & 2.37 & & &\\
        & 0.67,1.,1.  & 2.96 & 2.34 & -87.23 & 
0.63,0.91,1.22 & 2.66 & 2.33 & -89.51 & -\\
$2^+,0$ &             &      & 2.34 &   2.57 & 
            &      & 2.34 & 3.36 & 4.44\\
$4^+,0$ &             &      & 2.35 & 10.26 & 
            &      & 2.36 & 13.10& 14.08\\
$3^-,3$ & 0.72,1.,1. & 3.97 &  & &
0.64,1.13,1.20 & 3.84 & & &\\
        & 0.74,1.,1. & 2.96 & 2.54 & 8.57 & 
0.72,0.98,1.11 & 2.88 & 2.54 & 9.55 & 9.64\\
\hline 
 & \multicolumn{7}{c}{straight line} & & \\
\hline 
$0^+,0;l$ &  0.71, 1.,1. & 3.30 & 3.18 &  17.71 &
0.81,1.,0.80 & 3.28 & 3.21 & 16.27 & 7.65\\
\hline
\end{tabular}
\caption{\label{tbl2}{\scriptsize $^{12}C$ nucleus. As in Table 1,
    only results with  $T=0$ states are shown. See Table 1 for
    definitions and units.}}
\end{center}
}
\end{table}

Even tough the structure of $^{12}$C offers more possibilities of
forming clusters than $^8$Be, we will perform an analysis parallel to
this last one, that is, we will consider that the nucleons are
distributed around one center  or around three centers forming three
$\alpha$ particles. In Table 2, we show the results obtained for
$^{12}$C. When only one center is considered, we have used the
configuration $[(0,0,0)^4,(1,0,0)^4,(0,1,0)^4]$ as generating vector
of the rotational band of the ground state. This configuration is axially
deformed even for isotropic oscillators and
deformation of the oscillators only reforces this axial asymmetry. The
result with spherical harmonic oscillator agrees basically with the
one obtained by Dufour and Descouvemont \cite{DuDe-1996} who studied
this nucleus with this approximation and distributing the nucleons in
$\alpha$-clusters always with a spherical harmonic oscillator. When 
deformations in the harmonic oscillator potential are allowed, a
similar behavior to the one discussed for $^8$Be is found, with an important
increase of $10 \%$ in the ground state bound energy and a better
agreement in the excitation energy of the states in the rotational
band of the ground state  ($J^\pi=2^+,4^+$). We have also considered the
configuration $[(0,0,0)^4,(0,0,1)^4,(0,0,2)^4]$ as generating vector
to approximate the state $J^\pi=0^+$ with excitation energy of $7.65$
MeV. This state will be also described as three $\alpha$ particles
forming a straight segment on the $z$-axis. The results for this state
are very bad for the isotropic case, they improve for the deformed
case but the excitation energy provided is twice the experimental one
showing very important deviations compared to the ones found for the
rotational band of the ground state.

When the nucleons are grouped in three $\alpha$ particles, the
distribution forming triangles provides the lowest energy and can be
taken as generator of the rotational band of the ground state. Let us
begin discussing the results obtained for an equilateral triangle. In
this case, passing from isotropic to anisotropic oscillators provides
an improvement in the ground state energy of about $2.6 \%$. There is
also a better description of the excitation energy of the states in the rotational band
$J^\pi=2^+,4^+$. With respect to the state $J^\pi=3^-$ at $9.64$ MeV
both approximations provide quite similar results overestimating the
excitation energy. The results shown here are similar to the ones 
obtained by Dufour and Descouvemont \cite{DuDe-1996} for the isotropic
case and, for the deformed oscillator case fit quite well to the ones
obtained using GCM. It should be pointed out that the anisotropic
case, the three optimal strenghs are different. This can interpreted
as some polarization effects induced by the triangular distribution of
the clusters.

The possibility that the three $\alpha$ particles form an
isosceles triangle has been also investigated. In this case,  a new distance and a new
harmonic oscillator potential are added as additional variational parameters. 
These parameters are shown in the Table
2 in the following way: the first set of oscillator strenghts
corresponds to the potential experimented by two $\alpha$ particles
begin the first distance shown the distance between these two particles 
and the second set is the oscillator strenghts experimented by the
other $\alpha$ particle and the distance between this particle and the other
two. The results provide quite slight modifications in the ground state and in
the rotational band when they are compared to the equilateral triangle
case. However, there is a quite important effect on the
$J^\pi=3^-$ improving, greatly in the deformed case, the value of
the excitation energy.
Finally three $\alpha$ particles aligned are used to describe the
state  with $J^\pi=0^+$ and excitation energy of $7.65$ MeV. The
results obtained are even worse than the ones obtained for this state
with one center.

\begin{table}
{\scriptsize
\begin{center}
\begin{tabular}{|l|llcr|llcr|r|}
\hline
$J^{\pi},K$ & $(\alpha_x,d_y,d_z)$ & $\Delta$ & $\sqrt{<r^2>}$ &  $E$ &
 $(\alpha_x,d_y,d_z)$ & $\Delta$ & $\sqrt{<r^2>}$ & $E$ &\\
\hline
 & \multicolumn{7}{c}{(1):1-center} & &\\
\hline
$0^+,0$ & 0.61,1.,1. &   & 2.29 & -46.13 & 0.60,1.32,0.84 &   & 2.38 & -54.08
& -\\ 
$2^+,0$ &            &   & 2.29 &   1.96 &     &   & 2.39 &   3.23 & 3.37\\
$4^+,0$ &            &   & 2.29 &   6.65 &     &   & 2.44 &   14.17 & -\\
$2^+,2$ &            &   & 2.29 &   2.00 &     &   & 2.39 &   4.21 & 5.96\\
$3^+,2$ &            &   & 2.29 &   3.96 &     &   & 2.40 &   7.42 & -\\
$1^-,1$ & 0.59,1.,1. &   & 2.45 & 12.04 & 0.64,1.18,0.72 &  & 2.61 & 6.58 & 5.96\\
$2^-,1$ &            &   & 2.45 & 12.84 &                &      & 2.62 &  8.05 & 6.26\\
$3^-,1$ &            &   & 2.45 & 12.92 &                &      & 2.63 &  10.02 & 7.37\\
$0^+,0;l$ & 0.56,1.,1. & & 2.61 & 19.92 & 0.66,1.21,0.66 &  & 2.87 &  3.85 & 6.18\\
\hline 
 & \multicolumn{7}{c}{(2):3-centers $2\alpha+nn$} & &\\
\hline
$0^+,0$ & 0.72,1.,1. & 3.01 &      &        & 0.66,1.05,1.20 & 3.09 &  & &\\
        & 0.61,1.,1. & 2.60 & 2.38 & -56.00 & 0.48,1.19,1.66 & 2.51 & 2.43 & -58.78 & -\\
$2^+,0$ &            &      & 2.38 &   2.43 &                &      & 2.44 &   4.39 & 3.37\\
$4^+,0$ &            &      & 2.41 &  12.55 &               &      & 2.49 &  18.06 & -\\
$2^+,2$ &            &      & 2.39 &   5.74 &                &      & 2.44 &   5.67 & 5.96\\
$3^+,2$ &            &      & 2.40 &   8.06 &                &      & 2.46 &  10.26 & -\\
$1^-,1$ & 0.71,1.,1. & 3.68 &      &        & 0.63,1.15,1.23 & 3.21 & & &\\
        & 0.34,1.,1. & 3.38 & 3.04 &   9.50 & 0.26,1.11,1.57 & 4.75 & 3.29 &   9.89 & 5.96\\
$2^-,1$ &            &      & 3.04 &  11.47 &                &      & 3.30 &  12.33 & 6.26\\
$3^-,1$ &            &      & 3.08 &  14.05 &                &      & 3.39 &  15.46 & 7.37\\
$0^+,0:l$ &     0.67,1.,1.     & 2.48     &  2.90    &  13.76      & 0.66,1.19,0.70 & 1.65 & 2.84 &   8.12 & 6.18\\
\hline 
 & \multicolumn{7}{c}{(3):2-centers $^{6}He +\alpha$} & &\\
\hline
$0^+,0$ & 0.62,1.,1.  &      &      &        & 0.53,1.51,1.16 &  &  & & \\
        & 0.76,1.,1. & 3.02 & 2.40 & -55.67 & 0.79,0.99,0.84 & 3.02 & 2.44 & -59.23 & -\\
$2^+,0$ &              &     & 2.41 &   4.61 &    &                & 2.46 &  4.35 & 3.37\\
$4^+,0$ &              &     & 2.45 &  18.46 &    &                & 2.50 &  17.81 & -\\
$2^+,2$ &               &      & 2.40 &  2.71 &     & & 2.46 &   5.13 & 5.96\\
$3^+,2$ &               &      & 2.42 &  7.59 &     & & 2.47 &   9.67 & -\\
$0^+,0;l$ &  0.73,1.,1. &      &      &       & 0.62,1.43,1.03 &  & &  &\\
          &  0.62,1.,1. & 4.69 & 2.96 & 8.29  & 0.71,0.99,0.67 & 4.19 & 2.94 & 7.33 & 6.18\\
\hline 
 & \multicolumn{7}{c}{(4):2-centers $^{8}Be+nn$} & &\\
\hline
$0^+,0$ & 0.63,1.,1.  &  & & & 0.51,1.54,1.38 &  &  & & \\
        & 0.60,1.,1..  & 1.72 & 2.31 & -47.14 & 0.54,1.45,1.05 & 1.93 &
        2.40 & -56.14 & -\\
$2^+,0$ &            &   & 2.31 &   2.60 &           &  & 2.41 &  5.39 & 3.37\\
$4^+,0$ &            &   & 2.39 &  22.48 &           &  & 2.51 & 28.96 & -\\
$2^+,2$ &            &  & 2.31 &   2.07 &     &   & 2.41 &   3.84 & 5.96\\
$3^+,2$ &            &  & 2.31 &   4.71 &     &   & 2.43 &   9.34 & -\\
$0^+,0;l$ &  0.63,1.,1. &   & &  & 0.76,0.90,0.56 &  & &  &\\
          &  0.29,1.,1. & 2.57 & 3.00 & 11.65  & 0.76,0.90,0.59 & 1.11 & 2.86 & 5.83 & 6.18\\
\hline
\end{tabular}
\caption{\label{tbl3}{\scriptsize $^{10}Be$ nucleus. 
    Only results with  $T=1$ states are shown. You can read text for
    explanations about the 4 configurations and see Table 1 for
    definitions and units.}}
\end{center}
}
\end{table}

Let us study now the low energy states of $^{10}$Be. Different
distributions of the nucleons in clusters provide quite similar
energies for the ground state. That makes these configurations
possible candidates for describing the first states of this nucleus. 
We show in Table 3 the results obtained for 4 different distributions
in clusters using both spherical and deformed oscillators. 
These distributions are: (1) all the nucleons grouped around one center in
the configuration $[(0,0,0)^4,(0,0,1)^4,(0,1,0)^2]$; (2) the nucleons
distributed around three centers forming an isosceles triangles, two
centers with one $\alpha$ particle each and the third one with two
neutrons; (3) the nucleons distributed around two centers separated in
the $z$-axis, one with one $\alpha$ particle and the other one with 
the rest of nucleons in the configuration  
$[(0,0,0)^4,(1,0,0)^2]$ ($^6$He) and; (4) again the previous two
centers separated in the $z$-axis but
now one with two neutrons and the other eight nucleons in the configuration 
$[(0,0,0)^4,(1,0,0)^4]$ ($^8$Be). All distributions except (1)
require two vectors of oscillator strenghts. The first one shown in
Table 3 corresponds to the oscillator potential felt by the two
$\alpha$ particles in (2), the $^6$He cluster in (3) and the $^8$Be in
(4) while the second one corresponds to the two neutrons in (2) and
(4) and to the $\alpha$ cluster in (3). The number of different 
distances, $\Delta$, shown in the Table is zero
for (1) since there is only one cluster, one for (3) and (4) since
there are two clusters and two for (2) since the three clusters form an
isosceles triangle. The first distance shown in Table 3 corresponds to the
distance between the two $\alpha$ clusters while the second one is
the distance from the two neutrons cluster to every of the two $\alpha$ clusters.  
 
The ground state in all the distributions and approximations corresponds to the
state $J^\pi=0^+$ in the band $K=0$. We can see that the non-spherical
approximation (on the right) always provides a lower energy for all
the configurations compared to the corresponding spherical
approximation (on the left). It is remarkable to mention that all the
distributions used provide an intrinsically deformed state even in the
spherical case. In those distributions  where the intrinsic deformation is
important for the spherical case, the decrease of the energy gained by
the introduction of deformed oscillators is smaller. So the cases (2)
and (3), corresponding to three centers and two centers with $^6$He and
$\alpha$, are distributions intrinsically quite deformed and the
decrease of energy is only around 3 and 4 MeV, respectively. On the contrary,
the cases (1) and (4), corresponding to one center and two centers
with $^8$Be and $nn$, are not so deformed and the decrease is around 8
and 9 MeV. 
 
If we compare the energies obtained for different distributions when
the same approximation, spherical or non-spherical, is used, we can
see that the differences are not very important. The maximum
difference is around 10 MeV in the spherical case and around 5 MeV in
the non-spherical case, this corresponds to the $15 \%$ and $8 \%$ of
the total energy. The best energy is obtained for distribution (3) for
the non-spherical approximation but the difference to non-spherical distribution (2)
is only around $0.5$ MeV while this last distribution provides the lowest
spherical approximation energy. These small differences in the
energies indicate that all the proposed distributions are
good candidates for describing the ground state and there should be an
important linear dependence among them. The obtained energies for the
states $J^\pi=2^+,4^+$ in the band $K=0$, for the different
configurations and using the variational parameters used to minimize
the ground state energy, depend on that configuration specially when
spherical and non-spherical approximations are compared. However, the
energies provided for the non-spherical approximation are quite
similar for the four configurations and provide a reasonable estimate
of the experimental value of the first excited state  $J^\pi=2^+$. 

In Table 3, we have also included the excitation energies obtained for
the first states in the band $K=2$ with positive parity and $K=1$ with
negative parity. These first states can be considered to approximate
some of the states in the experimental spectrum of $^{10}$Be. It is
also proposed for every distribution an approximation to the first
excited state $J^\pi=0^+$ of the experimental spectrum that will be
discussed later. In the band
$K=2$, we have used the same variational parameters than for the case
$K=0$ since there are no relevant changes if we try to fix these
parameters minimizing the state $J^\pi,K=2^+,2$. The excitation
energies obtained for all the distributions are quite similar, except 
distributions (1) and (4). It is remarkable that the excitation energy
obtained for the state $J^\pi=2^+$ is quite close to the experimental
energy of the second $J^\pi=2^+$ state.

Let us discuss now the band $K=1$, 
up to generate negative parity states in the distribution (1), one of
the neutrons in $(1,0,0)$ should be promoted to the state $(0,0,2)$ from the gound
state configuration. The distributions (3) and (4) provide quite high excitation energies
for the negative parity states in the band $K=1$ so the results have
not been shown in Table 3. This is also the case for distributions (1)
and (2) when the variational parameters obtained for the ground state
are used. The results shown in Table 3 correspond to the one obtained
after fixing the variational parameters by minimizing the state
$J^\pi,K=1^-,1$. The root mean squared radius obtained for these states is
quite different for the positive parity states from bands $K=0,2$. In
distribution (1) there are important differences between the spherical
and non-spherical approximations, while these differences almost
disappear for distribution (2). The obtained results for the
excitation energies in the band $K=1$ are, in all cases, greater than the experimental
ones, being the non spherical approximation for one center ones the
most adequate.  

Finally, results for an approximation for the first excited
$J^\pi=0^+$ state, experimentally $6.18$ MeV higher than the ground
state, are shown Table 3 for all the distributions as $0^+,0;l$. In the case of one
center, the configuration that moves the two neutrons from $(1,0,0)$ to 
$(0,0,2)$ from the ground state one is used. In distribution (2), the
three clusters are aligned in the $z$-axis with the two neutrons in
the middle position. In distribution (3), the $^6$He is in 
the configuration $[(0,0,0)^4,(0,0,1)^2]$ separated from the $\alpha$
in the $z$-axis and in the distribution
(4), the $^8$Be is in the configuration   $[(0,0,0)^4,(0,0,1)^4]$ and
the two neutrons separared in the $z$-axis. 
The results presented in the Table have the variational parameters
optimized to get the lowest excitation energy. Almost all the
configurations provide results quite close to the experimental ones.

The results obtained for the excited states with the three cluster
distribution, (2), are similar in general to the ones obtained using
GCM \cite{Desc-2002} with spherical harmonic oscillators and the same
Volkov interaction, although these authors fix the parameters in the
interaction for every parity. Our results also compare quite well to
the ones using AMD \cite{KHD-1999} who use a Volkov interaction that
includes three-body terms.     

\begin{figure}[h]
\begin{center}
\includegraphics[scale=0.57]{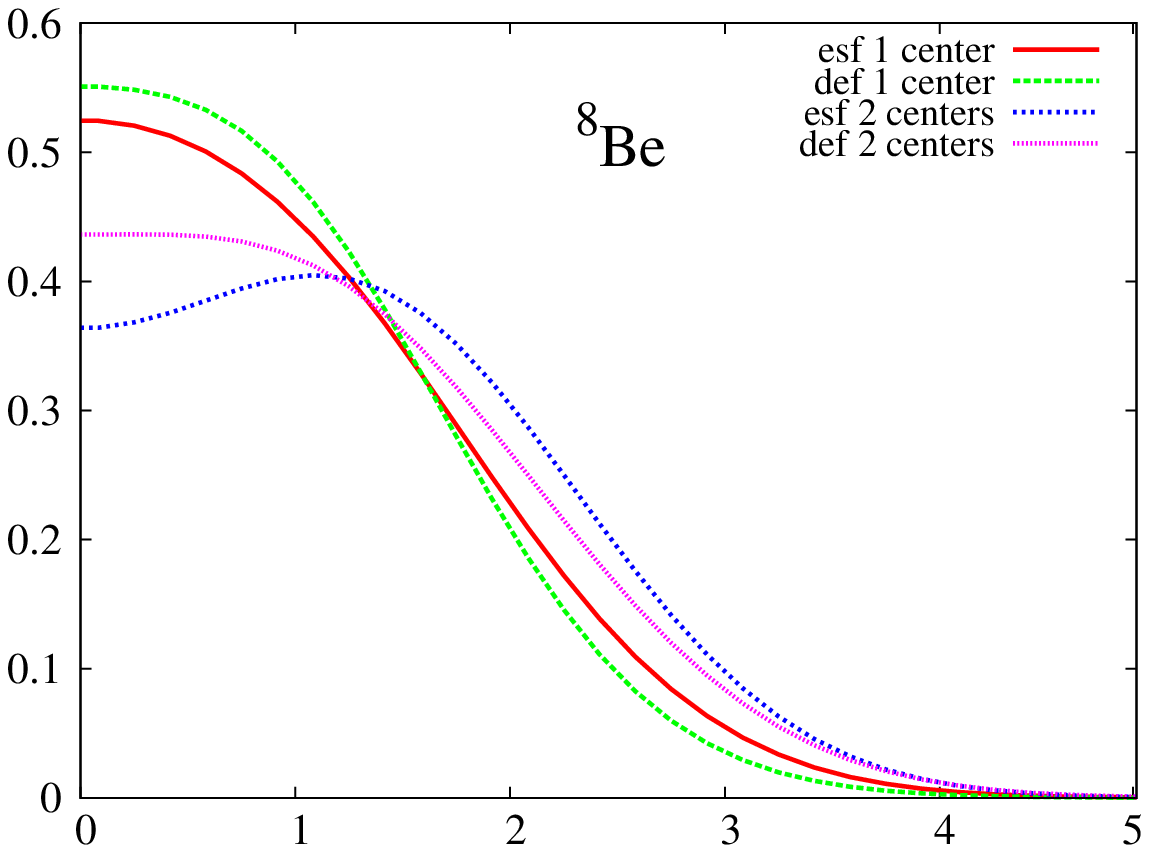}
\includegraphics[scale=0.57]{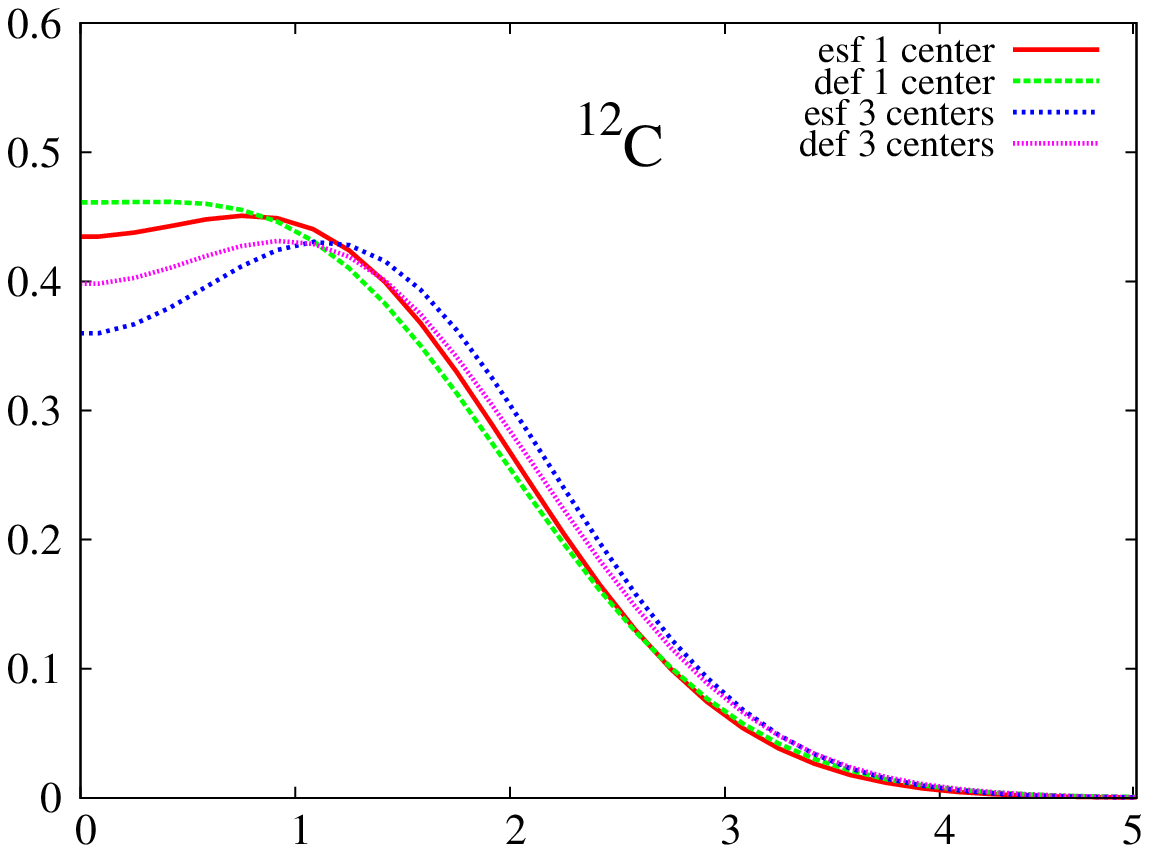}
\caption{\label{fig1} {\scriptsize Radial one-body density for $^8Be$
    (left) and $^{12}C$ (right) obtained for one-center and
    $\alpha$-clusters and for spherical and non-spherical approximations.}}   
\end{center}
\end{figure}

\section{Spatial density.}

A more straight image of the modifications caused by the use of
non-spherical harmonic oscillator functions for describing the
movement of the nucleons can be provided by simply comparing the
one-body density of the different distributions and approximations
previosly presented. An average image of the spatial distribution of
the nucleons comes from the radial one-body density, obtained after
integrating in the angular degrees of freedom.  In Figure \ref{fig1},
this radial density is shown for the nuclei: $^8$Be (left) and
$^{12}$C (right) corresponding to the one-center and forming two and
three $\alpha$-clusters, respectively. The radial density obtained for
the two distributions of $^8$Be are quite different, specially the two
$\alpha$-clusters in spherical approximation that present a
depression around the center of mass. In both distributions, spherical
approximation provides a more diffuse density than the non-spherical
one. For $^{12}$C the same behavior appears; however, for this
nucleus, all the densities present a depression around the center of
mass. This depression is more important for the three
$\alpha$-clusters distribution in spherical approximation and almost
dissapears for non-spherical one-center distribution. 

\begin{figure}[h]
\begin{center}
\includegraphics[scale=0.57]{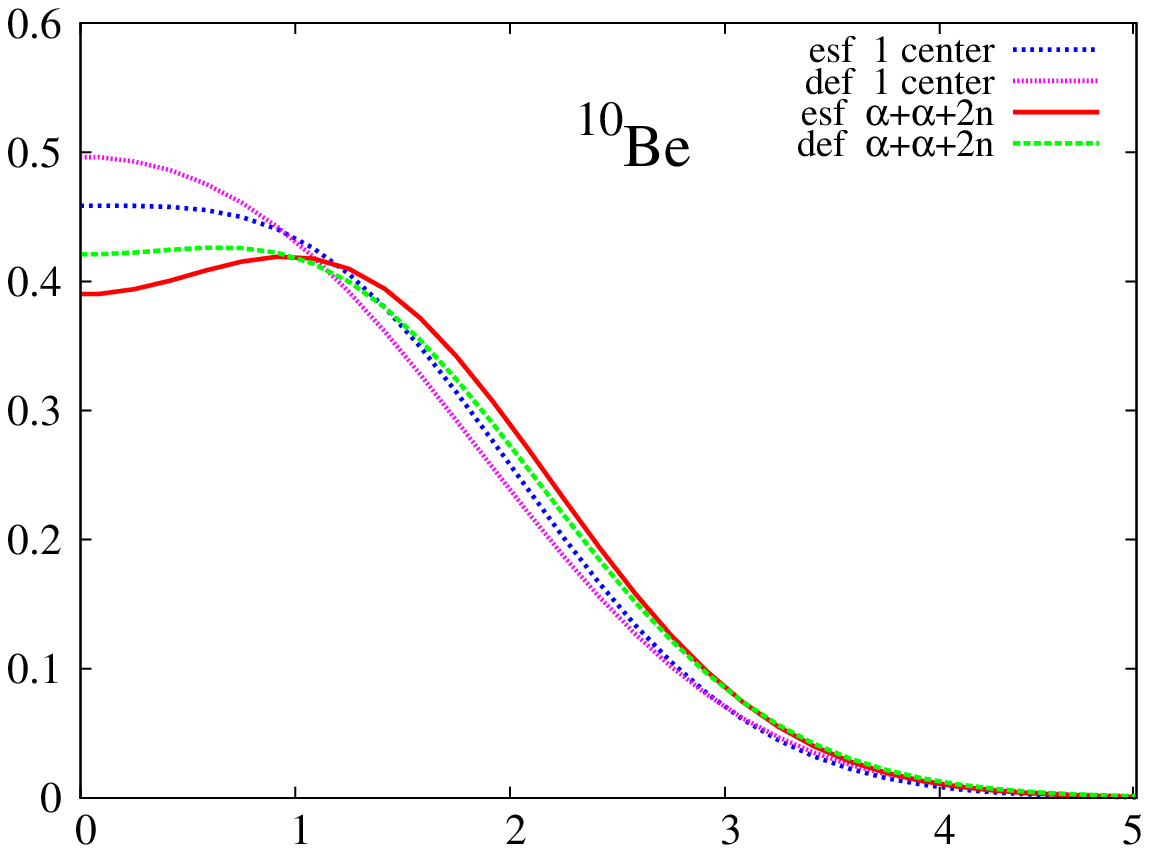}
\includegraphics[scale=0.57]{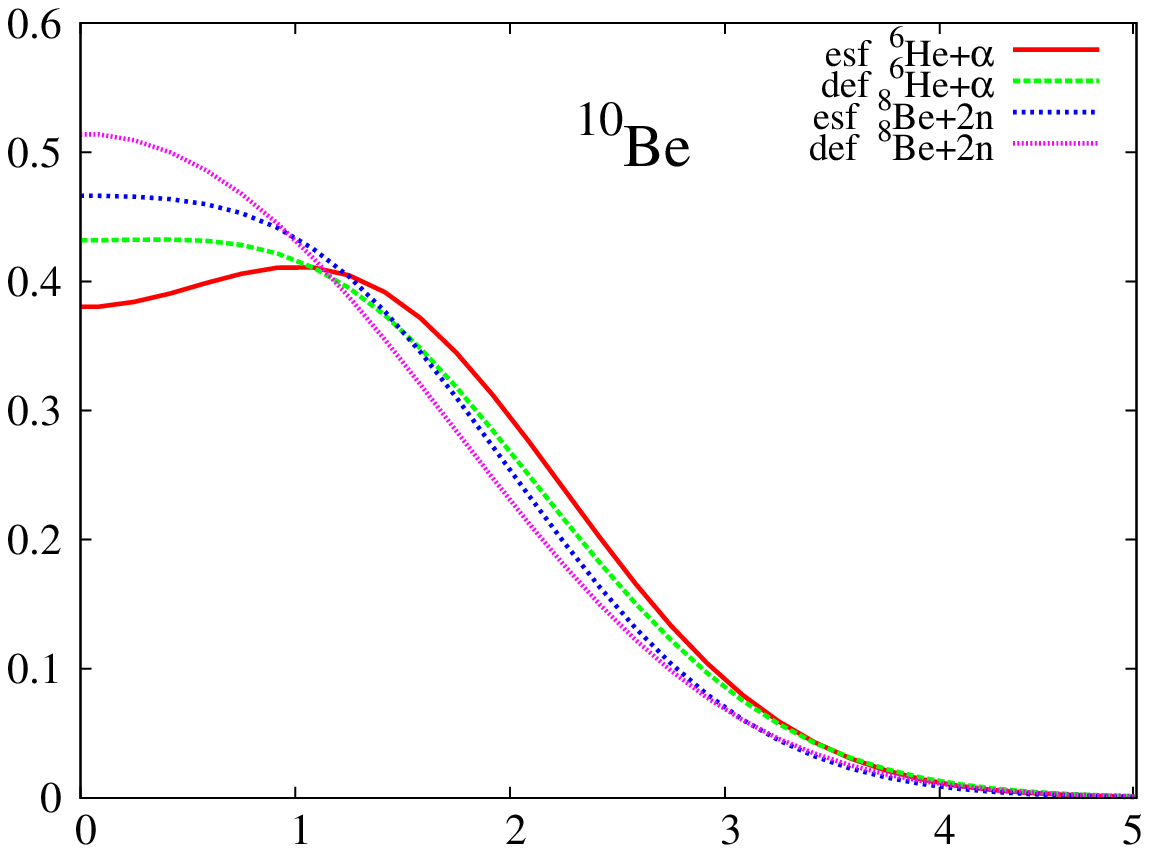}
\caption{\label{fig2} {\scriptsize Radial one-body density for  $^{10}Be$
    obtained for the four distribution studied: (1) and (2) on the
    left graph and (3) and (4) on the right graph. Every of the
    distributions is shown both within spherical
    and non-spherical approximations.}}
\end{center}
\end{figure}

The radial one-body density for the proposed distributions for
$^{10}$Be is shown in Figure \ref{fig2}. The behavior of the densities
is roughly similar to the one discussed for the two previous nuclei,
i.e., the spherical approximation provides more diffuse densities than
the non-spherical one for the four analyzed distributions. If we
compare the densities for the different distributions, we can see that
differences increase with the degree of clusterization, that is, when
we pass from one center, (1) to three centers (2) throught two
centers: (3) and (4), in this order.

\begin{figure}[t]
\begin{center}
\includegraphics[scale=0.28]{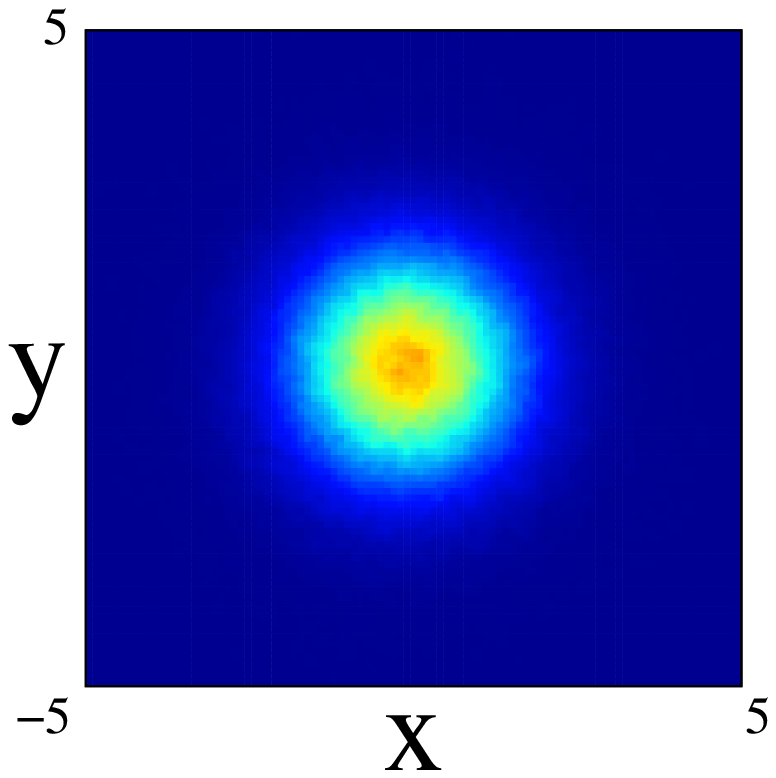}
\includegraphics[scale=0.28]{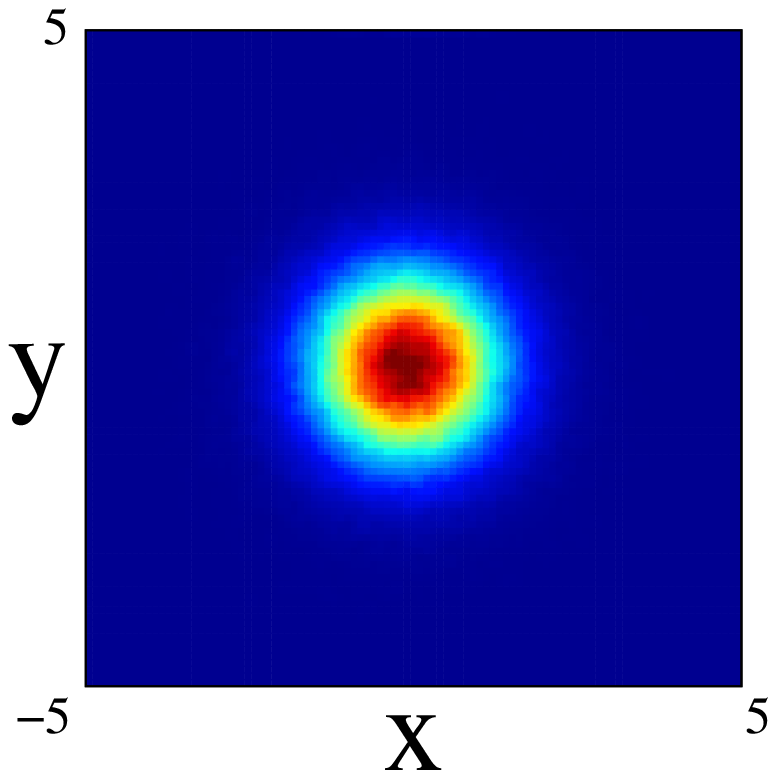}
\includegraphics[scale=0.28]{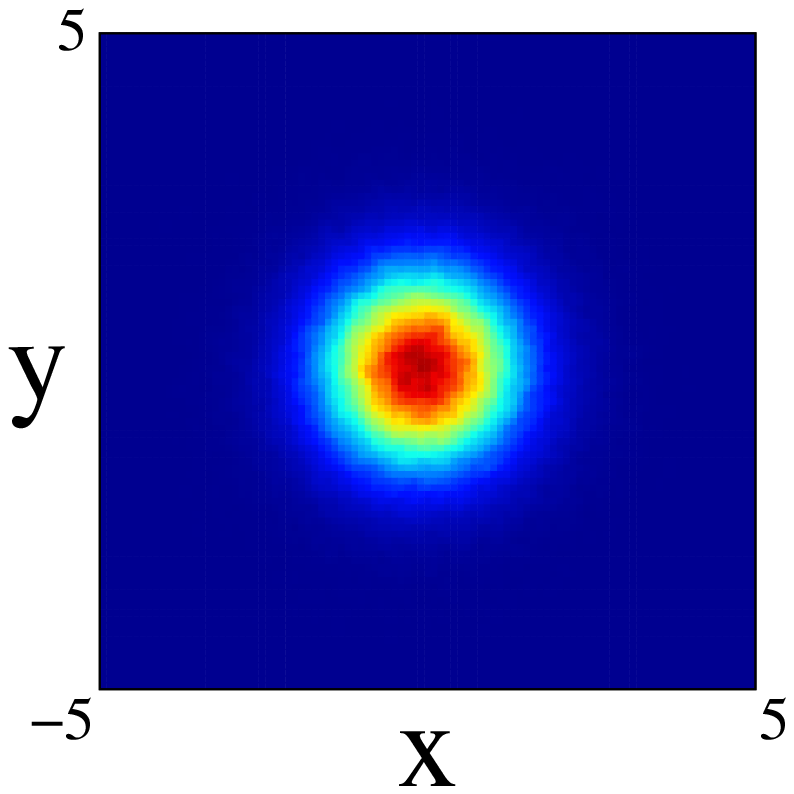}
\includegraphics[scale=0.28]{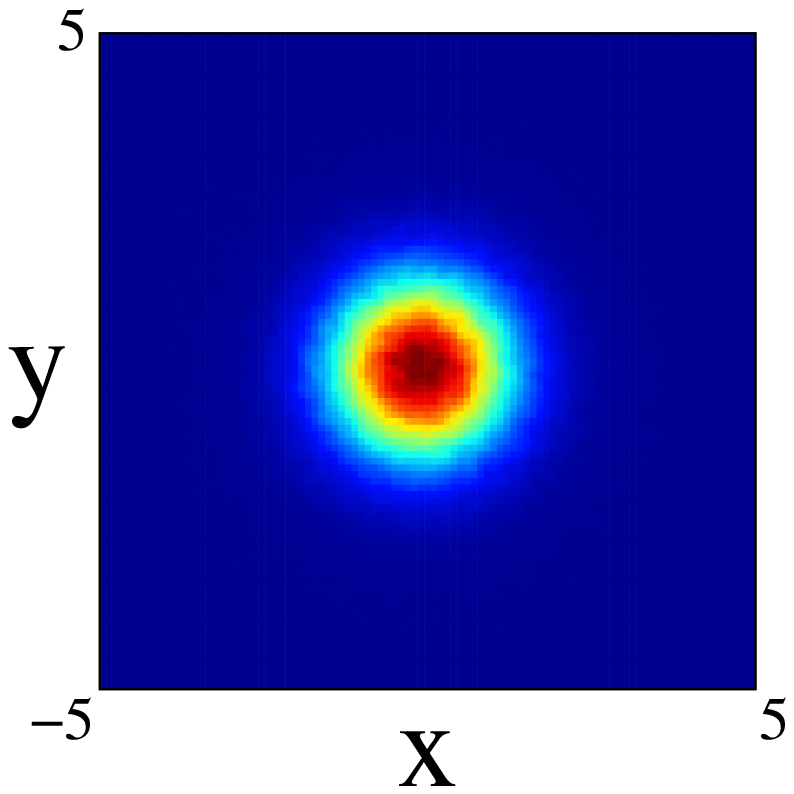}
\includegraphics[scale=0.28]{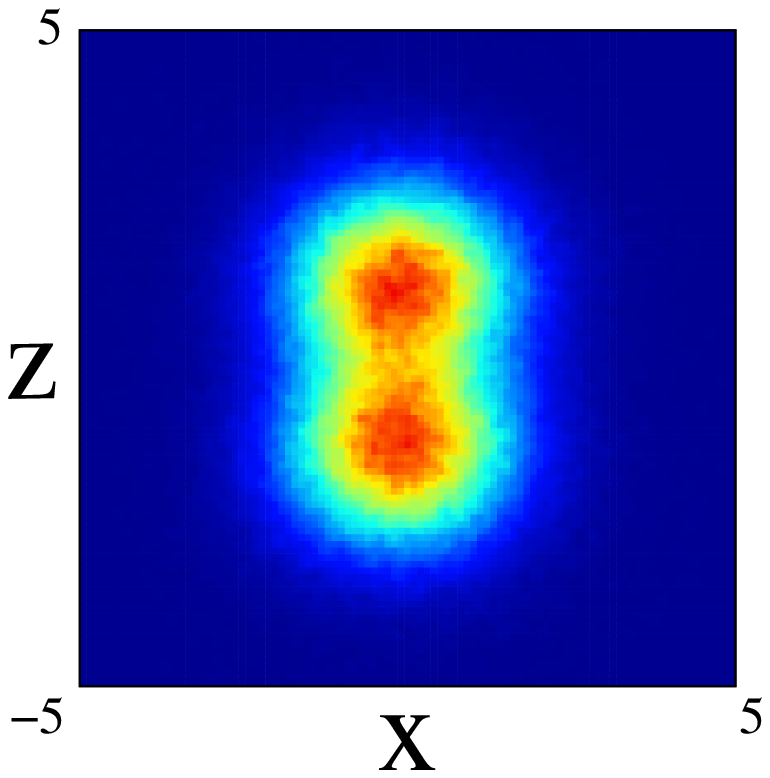}
\includegraphics[scale=0.28]{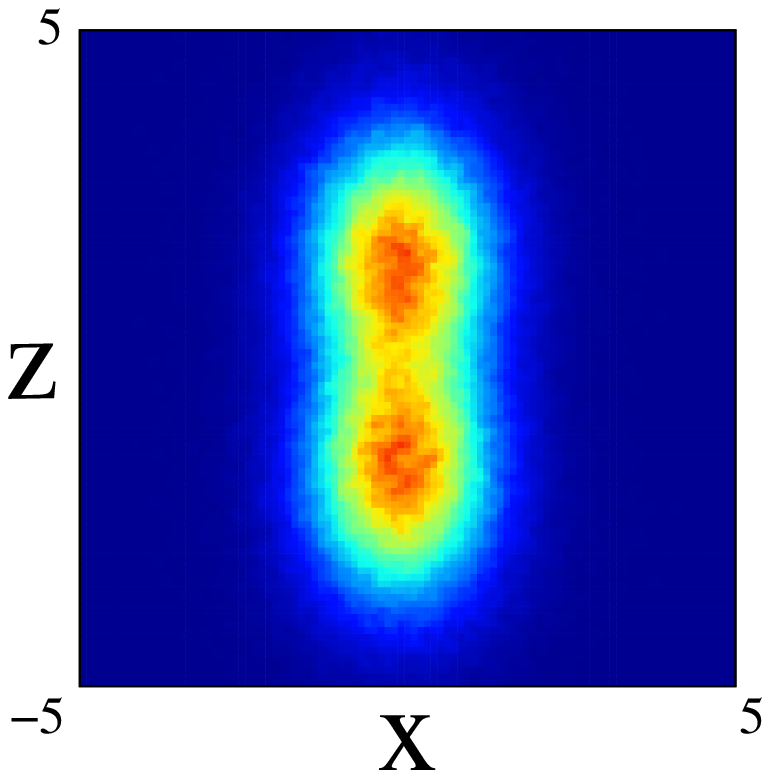}
\includegraphics[scale=0.28]{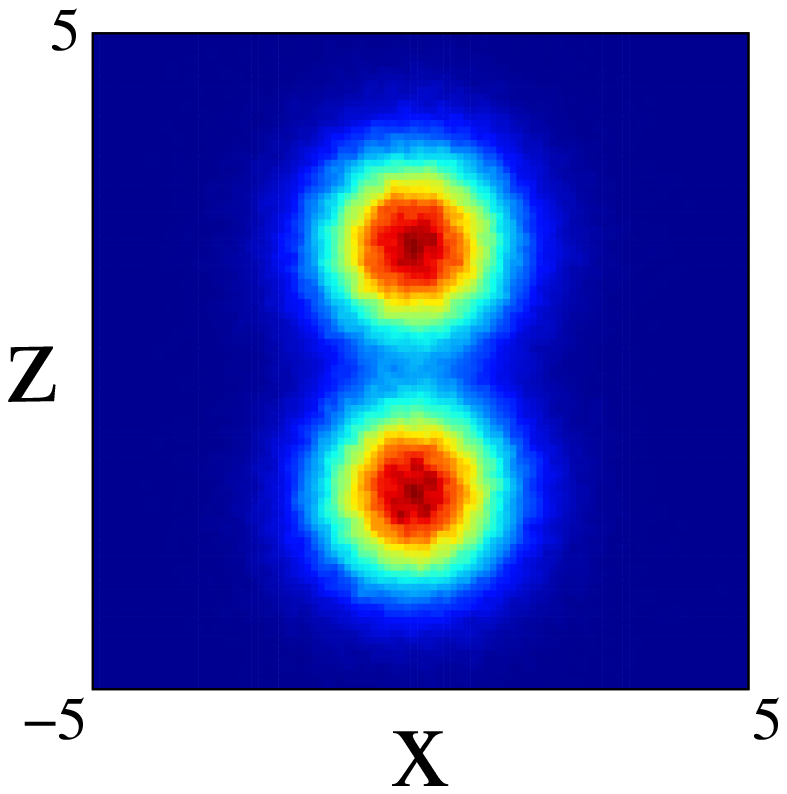}
\includegraphics[scale=0.28]{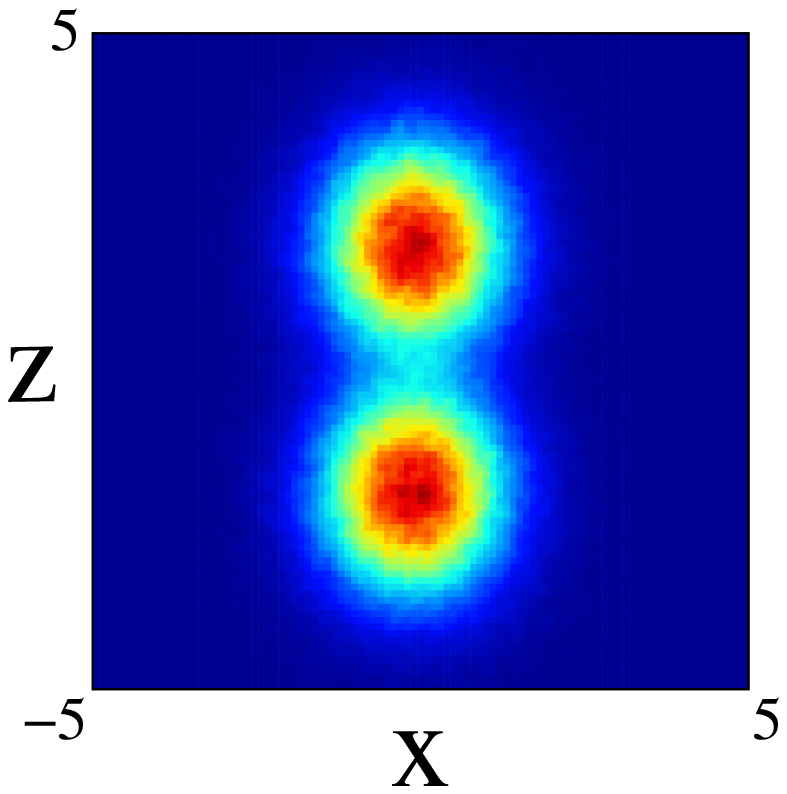}
\caption{\label{fig3} {\scriptsize $^8Be$ nucleus: Projection on the two
    cartesian planes: $xy$ (first row) and $xz$ (second row), since there
    is axial symmetry in this case,  of the density obtained for the one-center
    distribution (first two columns) and the two $\alpha$-clusters
    distribution (last two columns).  For every distribution, the left
    column is within spherical approximation and the right one is within
    the non-spherical.}}
\end{center}
\end{figure}

The angular average carried out for getting the radial one-body
density does not allow to explore the spatial distributions of the
nucleons since its intrinsic form is quite anisotropic for the studied
distributions. A more adequate image is obtained if the projections of
the spatial density on the three different cartesian planes is
performed. This projections have been obtained by a Monte Carlo
sampling of $10^6$ movements with an acceptance of $60 \%$ for all the
cases. For $^8$Be nucleus and due to the axial symmety around the
$z$-axis, the Figure \ref{fig3} shows the projection on the $xy$ plane
(first row) and on the $xz$ plane (second row) of the two
distributions studied and with spherical and non-spherical
approximations (see figure caption for further details). In the
projections on the $xy$ plane, we can see that the spherical
approximation presents a more diffuse proyection  than the
non-spherical one for the two distributions. However, in the
projection on $xz$, we can see that along the $z$-axis, the
diffusivity is more important in the non-spherical approximation
compared to the spherical one and the behavior reverses on the
$x$-axis. These differences can not be appreciated in the averaged
density since they compensate. In the two clusters distribution, an
slight overlap between the clusters appears and it is more important
for the non-spherical approximation. Finally, it is interesting to
note that the nucleus is structured in two centers along $z$-axis even in the case of
only one cluster, although these two centers are not so well defined in
the non-spherical approximation. 

\begin{figure}[t]
\begin{center}
\includegraphics[scale=0.28]{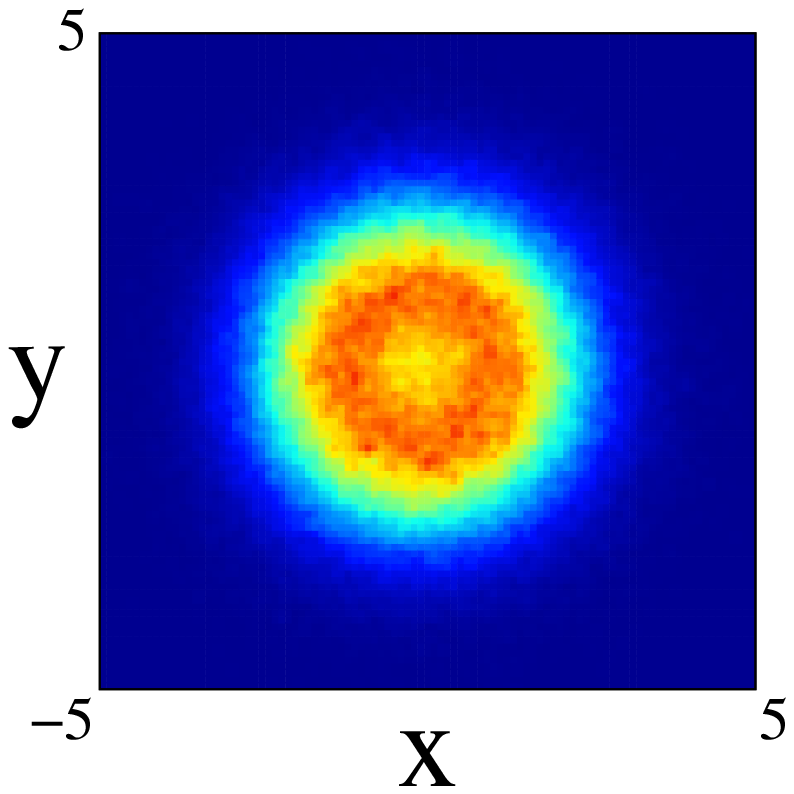}
\includegraphics[scale=0.28]{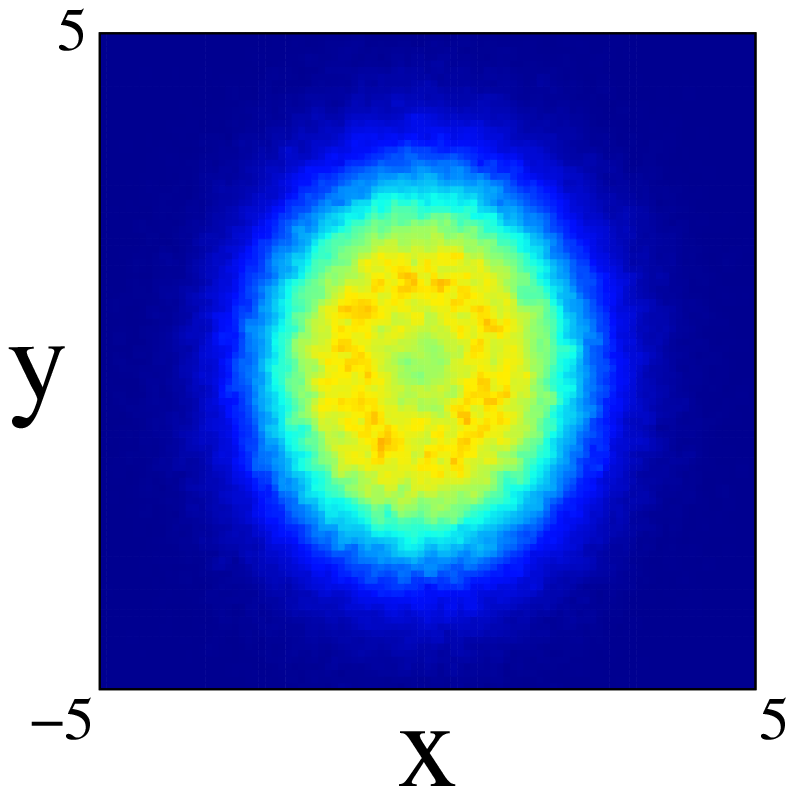}
\includegraphics[scale=0.28]{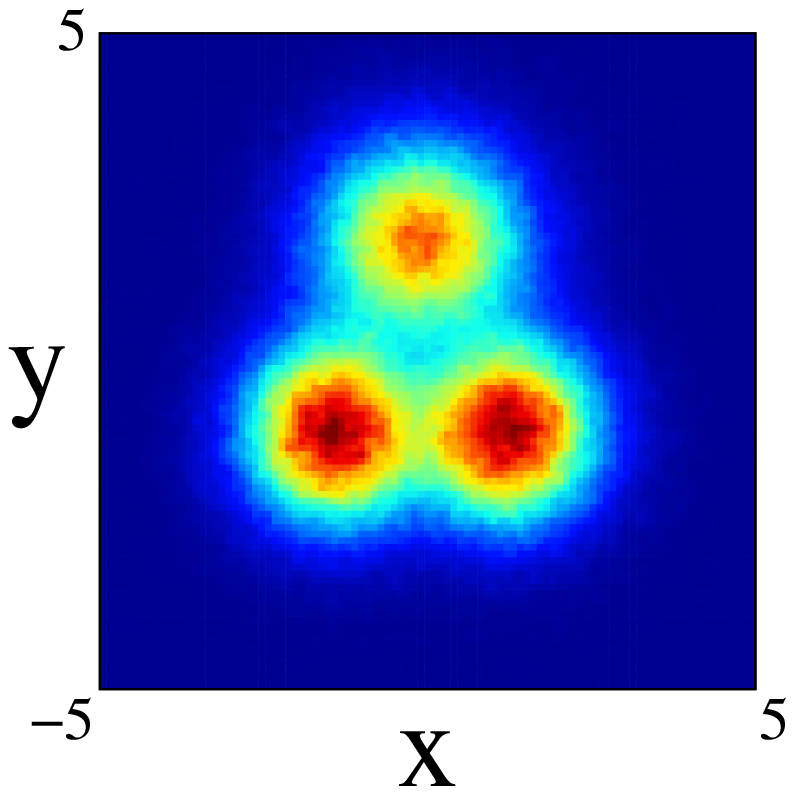}
\includegraphics[scale=0.28]{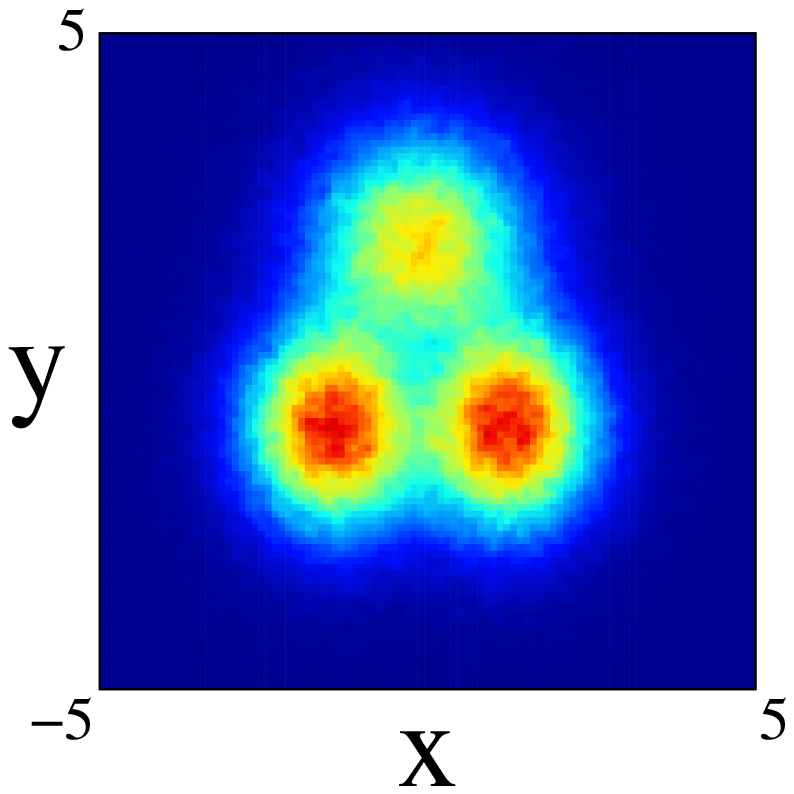}
\includegraphics[scale=0.28]{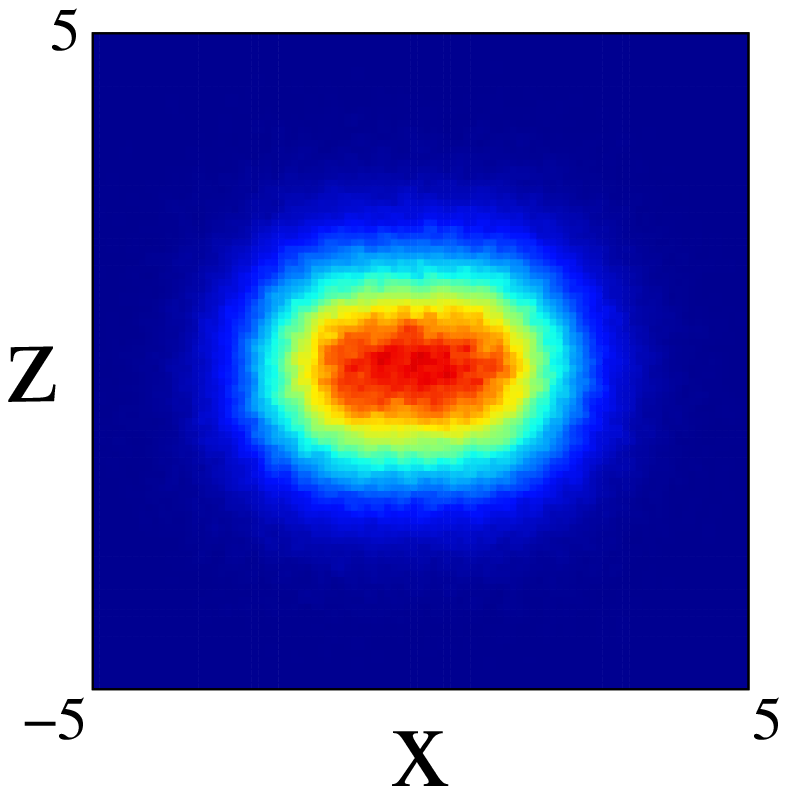}
\includegraphics[scale=0.28]{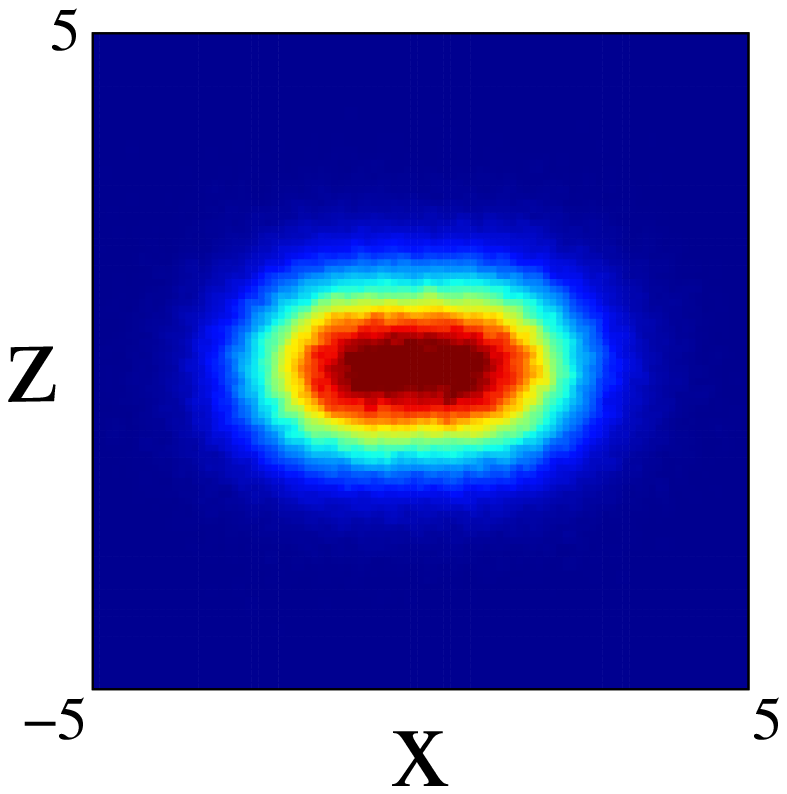}
\includegraphics[scale=0.28]{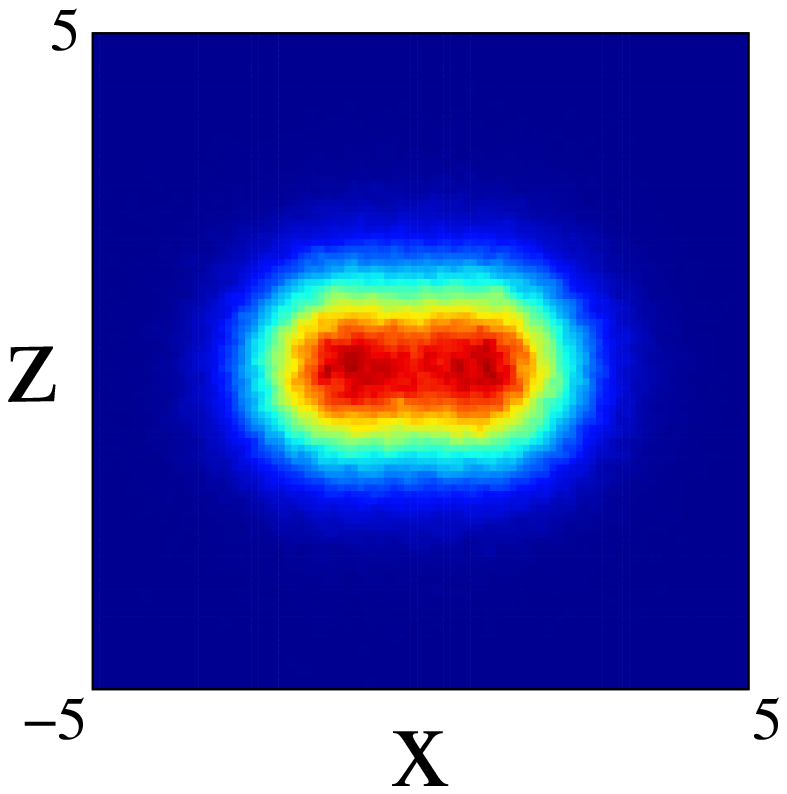}
\includegraphics[scale=0.28]{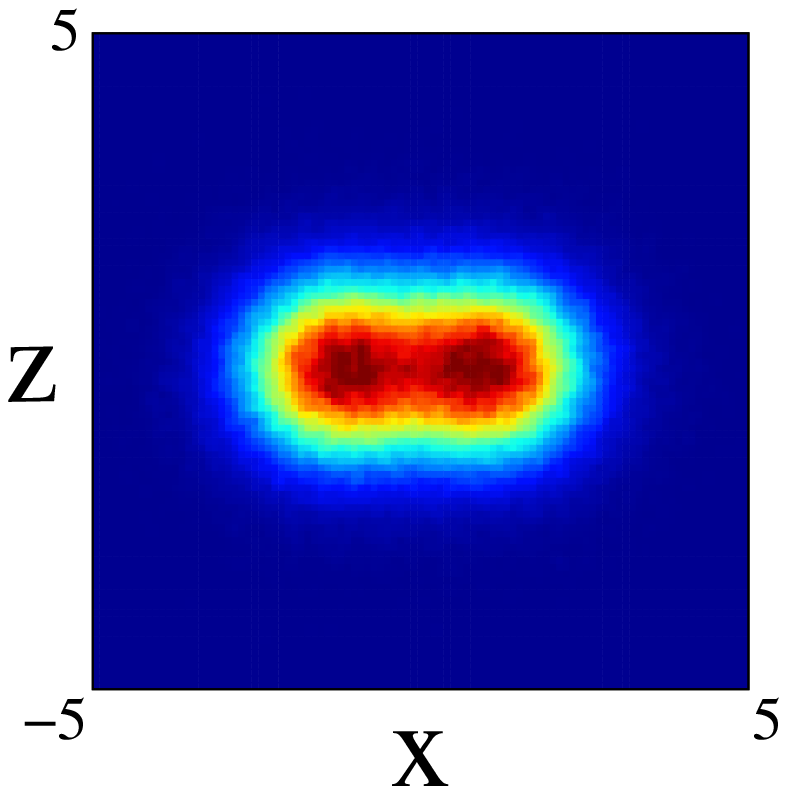}
\includegraphics[scale=0.28]{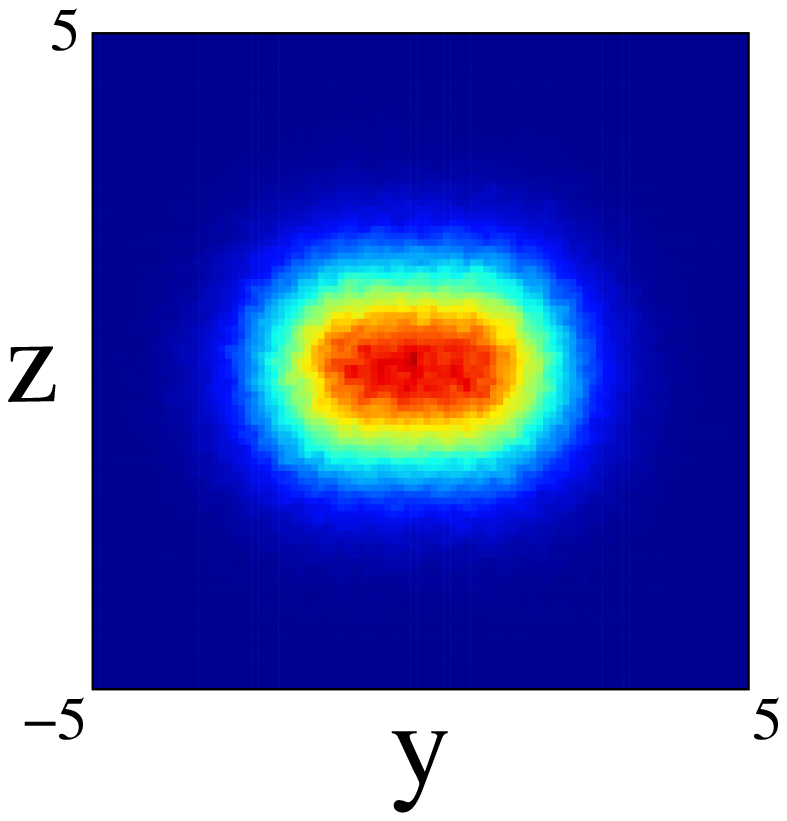}
\includegraphics[scale=0.28]{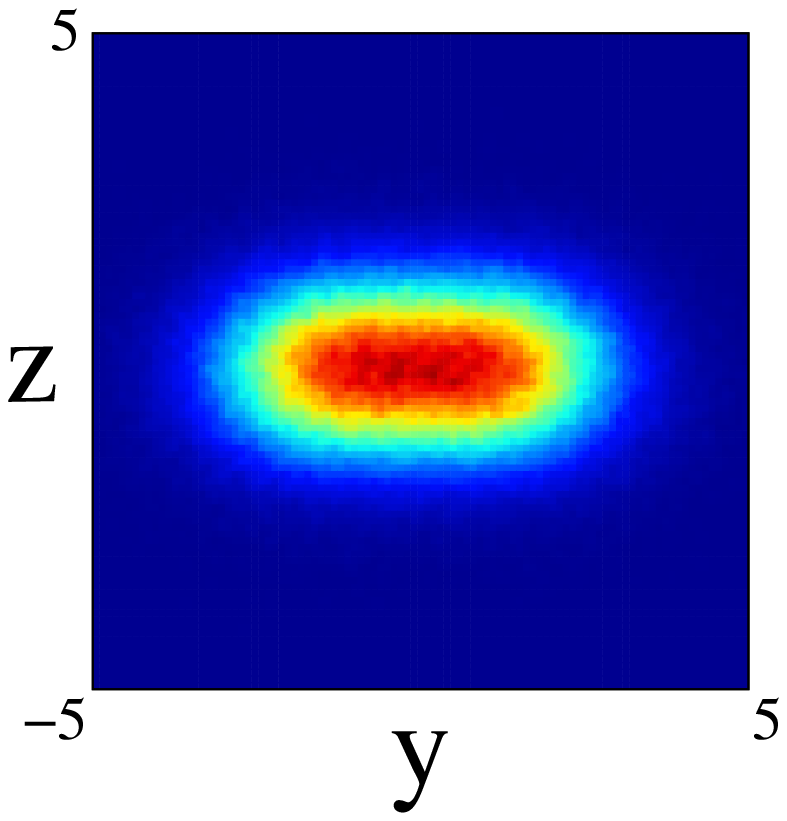}
\includegraphics[scale=0.28]{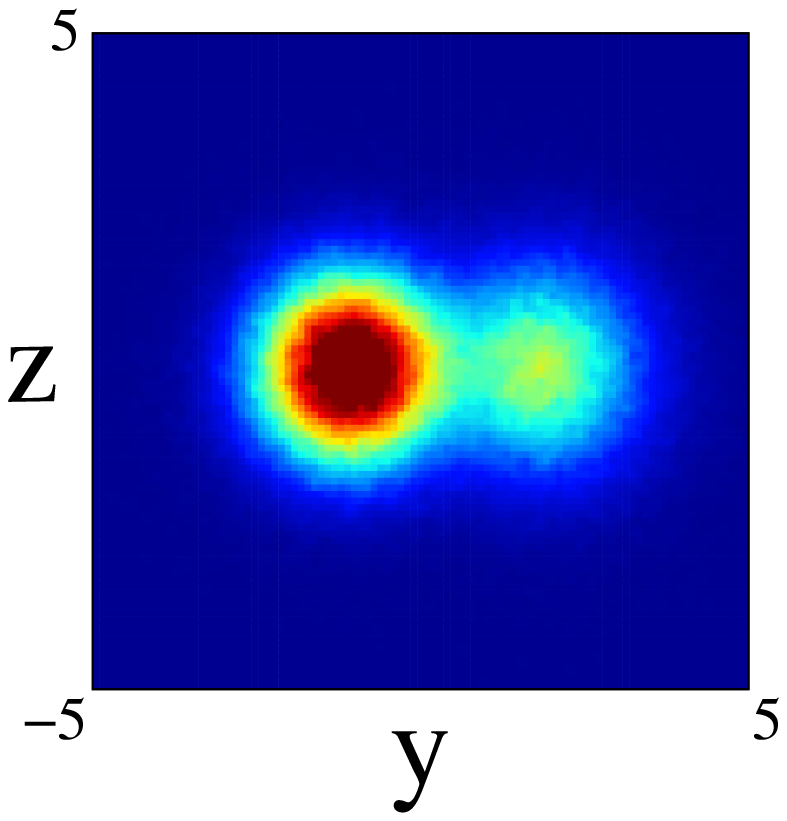}
\includegraphics[scale=0.28]{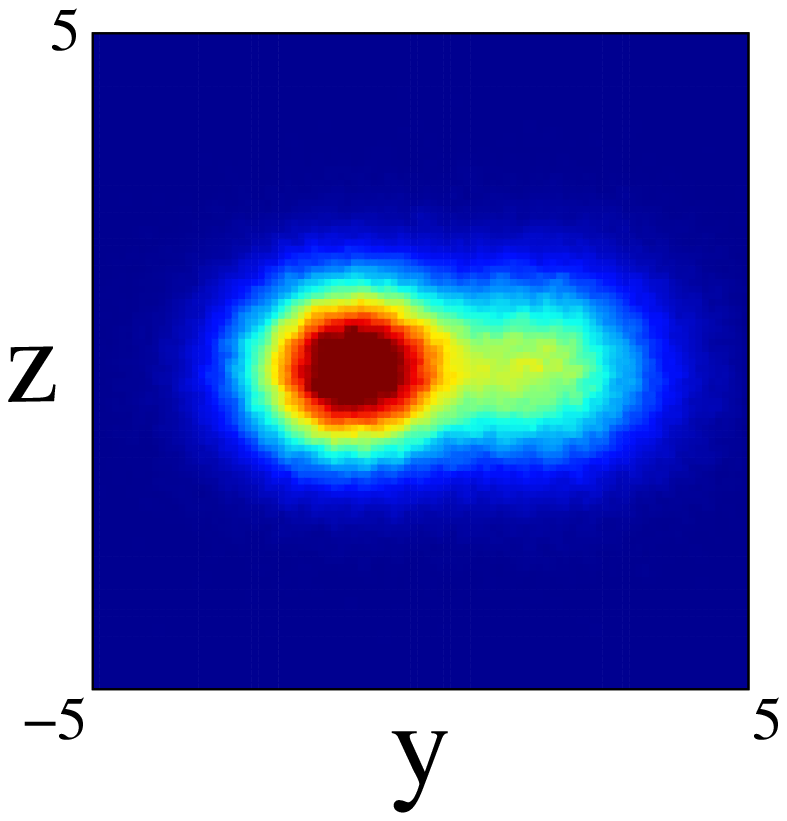}
\caption{\label{fig4} {\scriptsize $^{12}C$ nucleus:  Projection on the three
    cartesian planes: $xy$ (first row), $xz$ (second row) and $yz$
    (third row) of the density obtained for the one-center
    distribution (first two columns) and the three $\alpha$-clusters
    forming an isosceles triangle distribution (last two columns).
    For every distribution, the left column
    is within spherical approximation and the right one is within
    the non-spherical.}}
\end{center}
\end{figure}
 
In Figure \ref{fig4}, we show for $^{12}$C nucleus, the projections on
the three cartesian planes: $xy$ (first row), $xz$ (second row) and
$yz$ (third row) for the one center and three $\alpha$ clusters
forming an isosceles triangle distribution within spherical and
non-spherical approximations (see caption for more explanations). 
For both distributions, we can see important difference between the
two approximations. These are specially
important in the $xy$ plane for the one center case. For the three
clusters distribution (whose centers define the $xy$ plane) we can see
the loss of the equilateral symmetry on the different projections on
the $xz$ and $yz$ planes. This is also shown in the $xy$ plane seeing 
the difference of diffusion of one of the centers (the one on the
$y$-axis) relative to the other two. Finally, to remark that the
differences in the location of the nucleons are important as it
can be infered from Figure \ref{fig3}, even though the radial
density of both distributions is quite similar.

\begin{figure}[t]
\begin{center}
\includegraphics[scale=0.28]{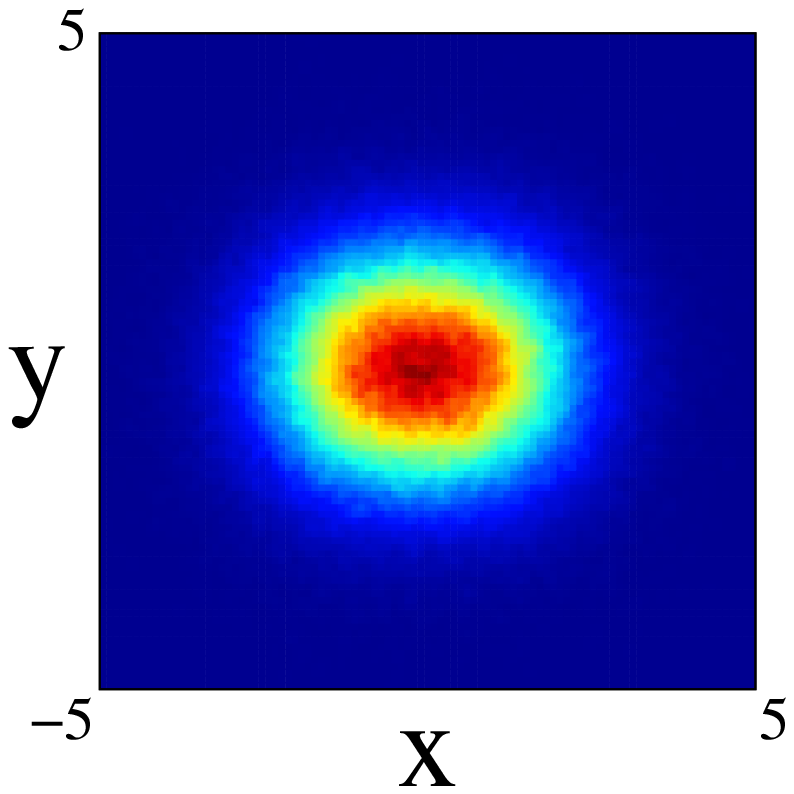}
\includegraphics[scale=0.28]{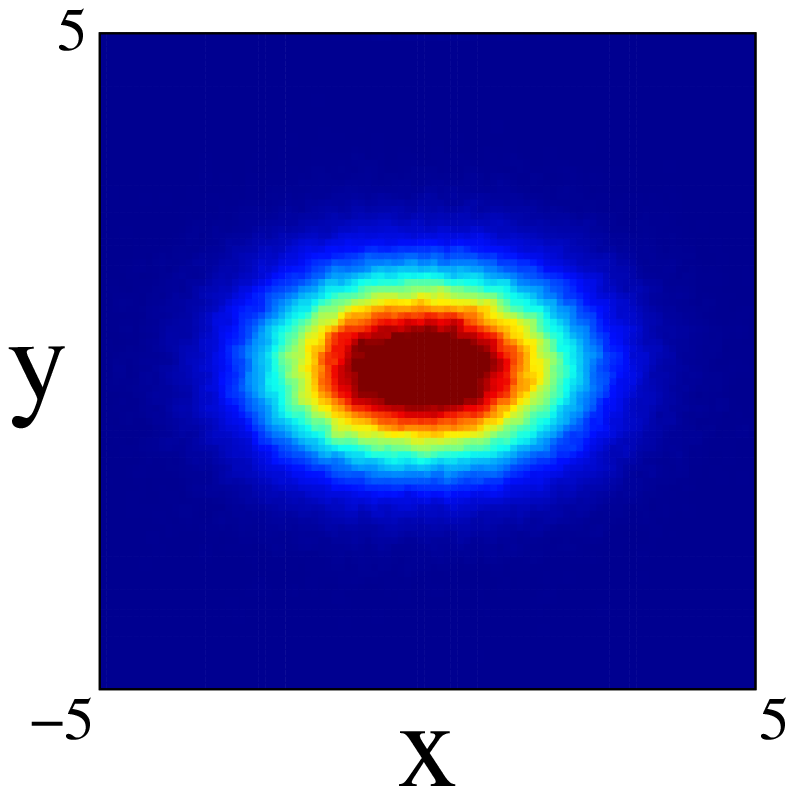}
\includegraphics[scale=0.28]{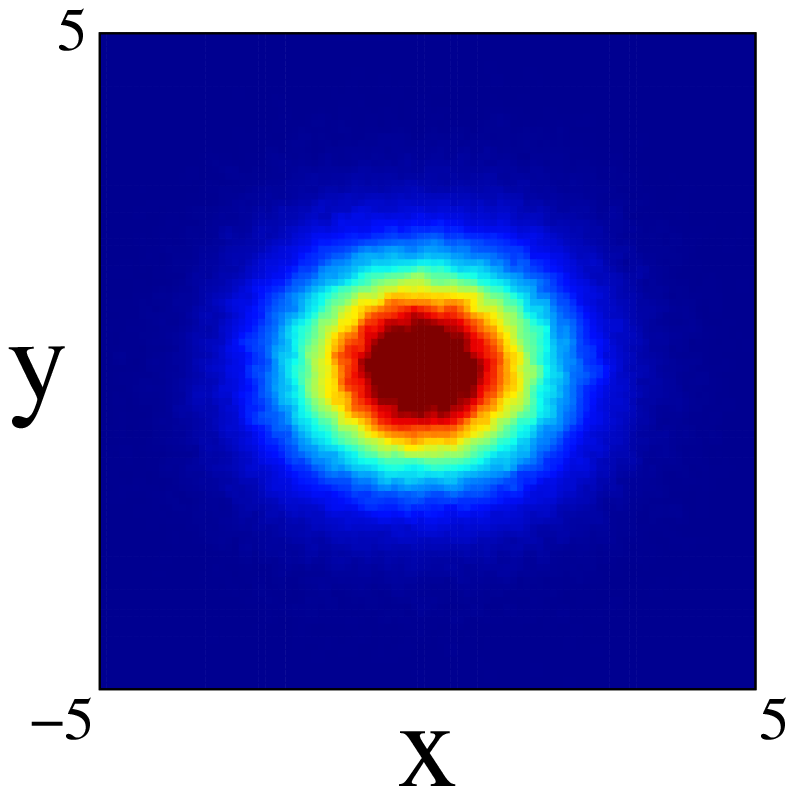}
\includegraphics[scale=0.28]{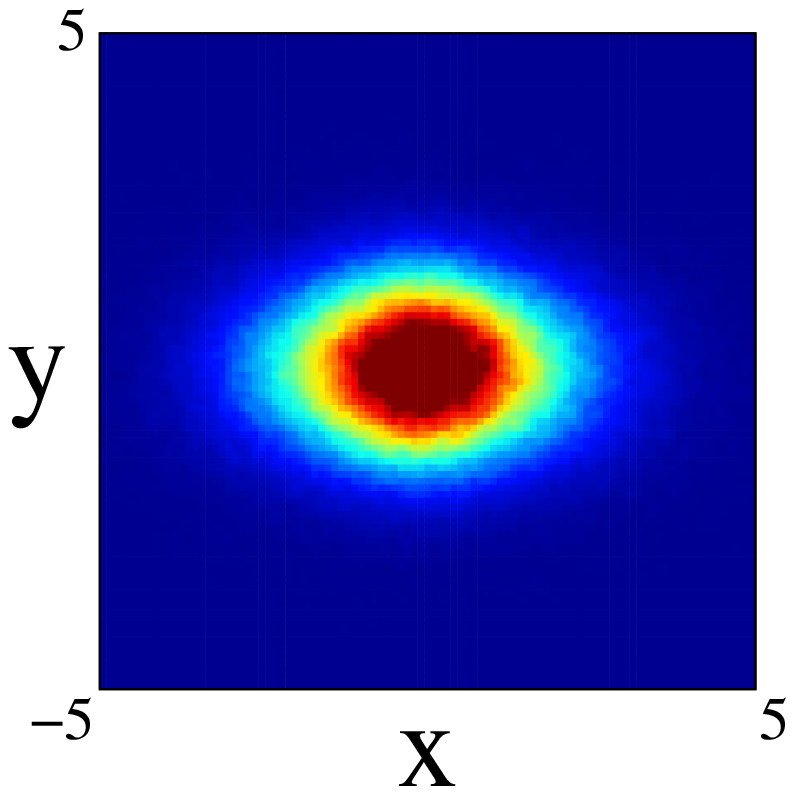}
\includegraphics[scale=0.28]{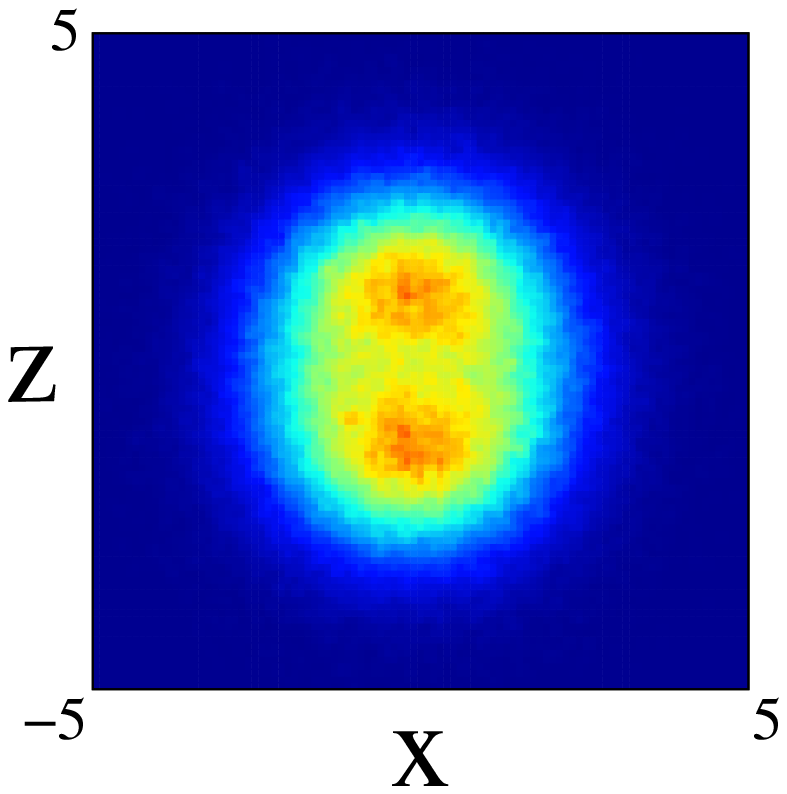}
\includegraphics[scale=0.28]{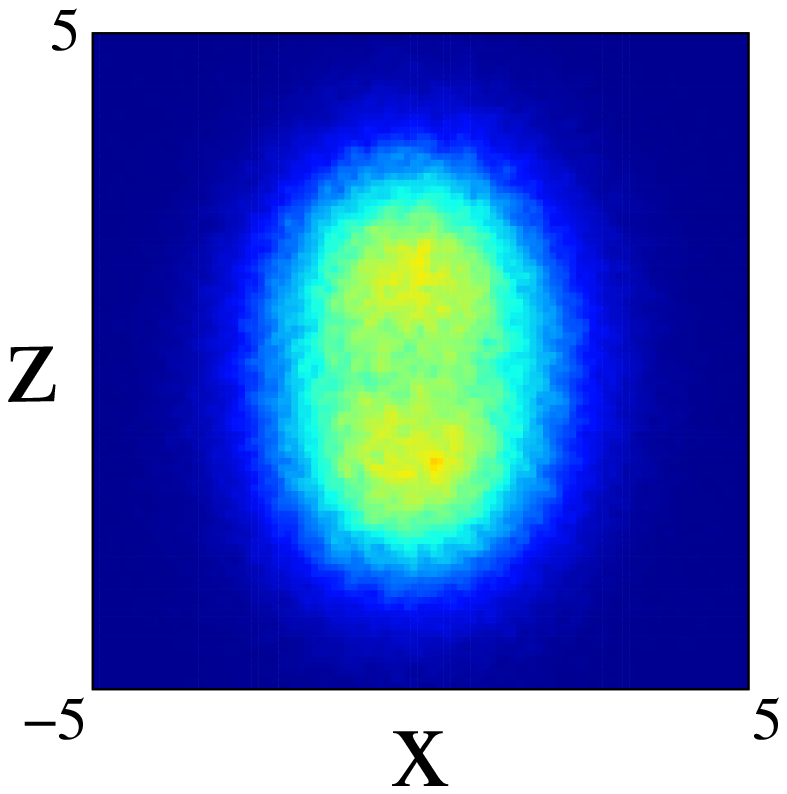}
\includegraphics[scale=0.28]{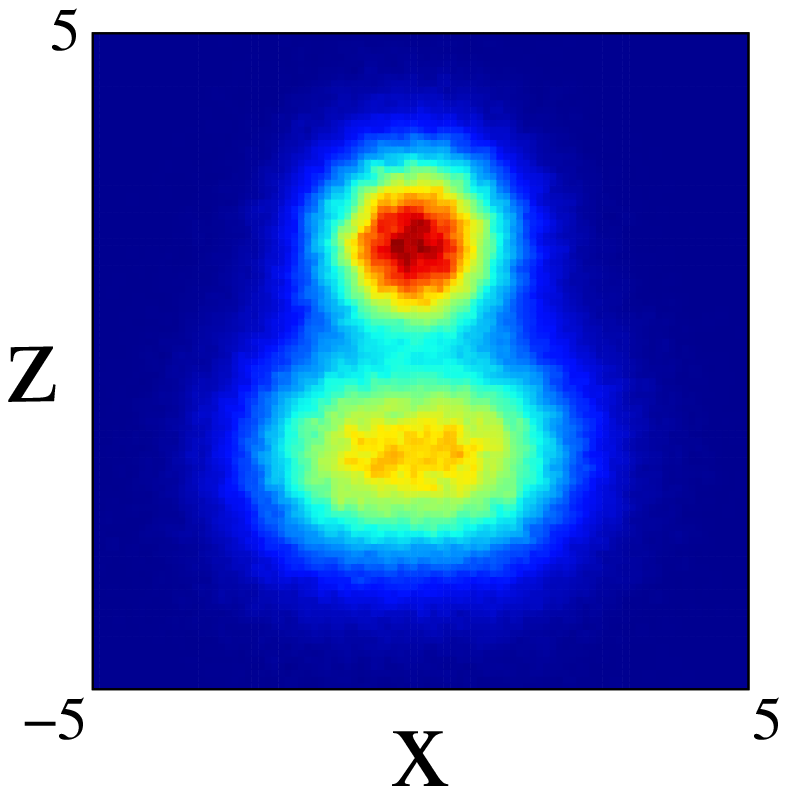}
\includegraphics[scale=0.28]{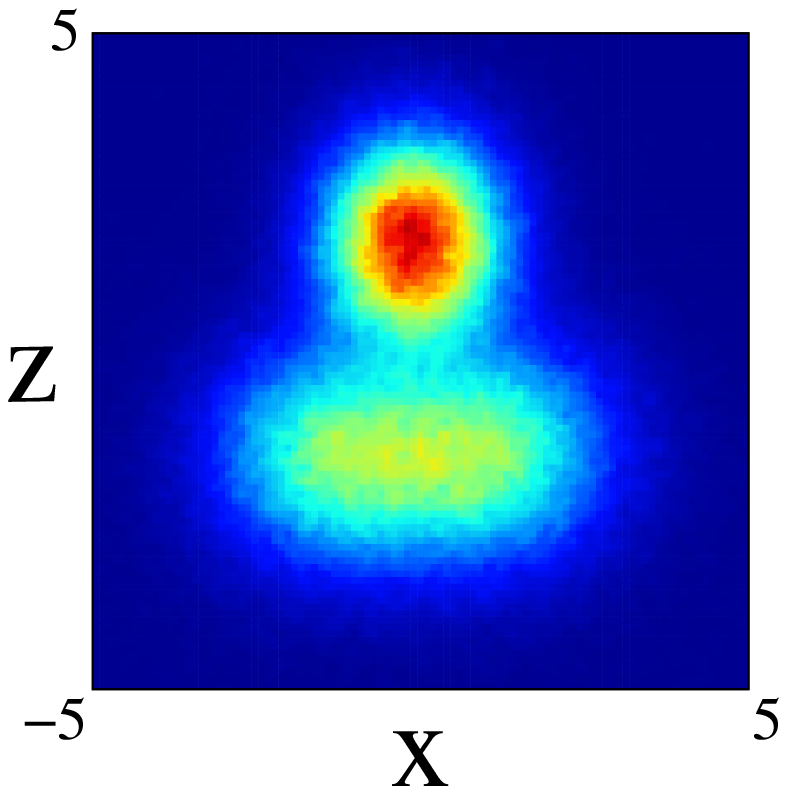}
\includegraphics[scale=0.28]{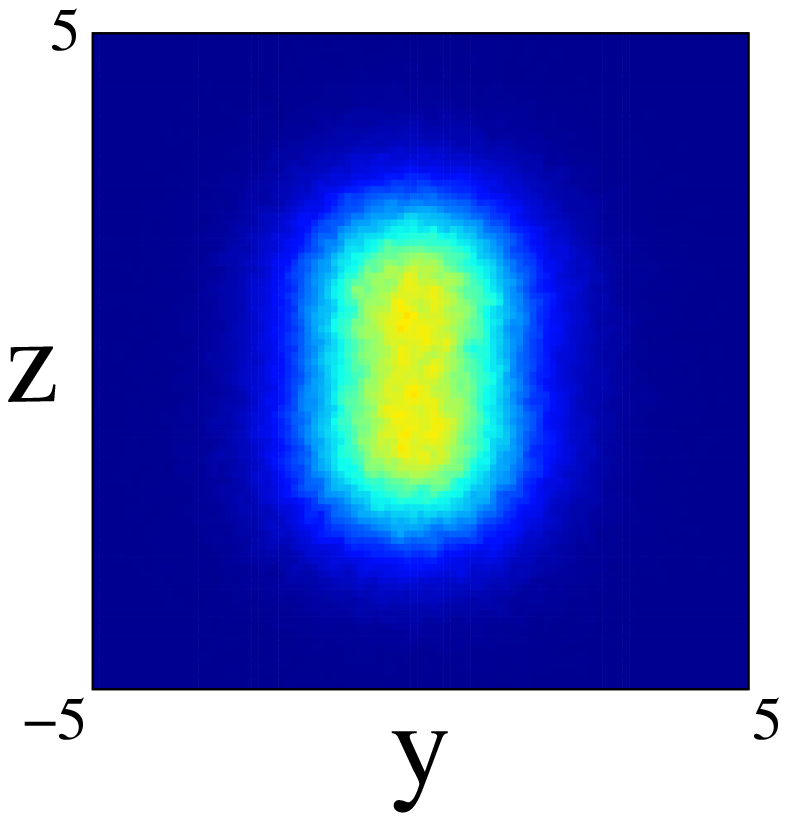}
\includegraphics[scale=0.28]{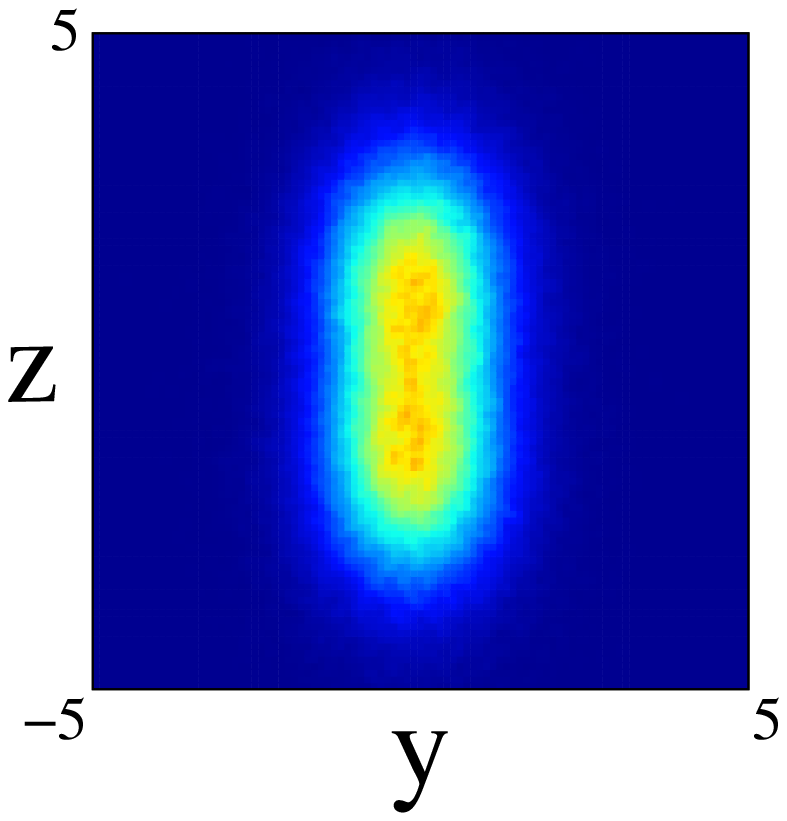}
\includegraphics[scale=0.28]{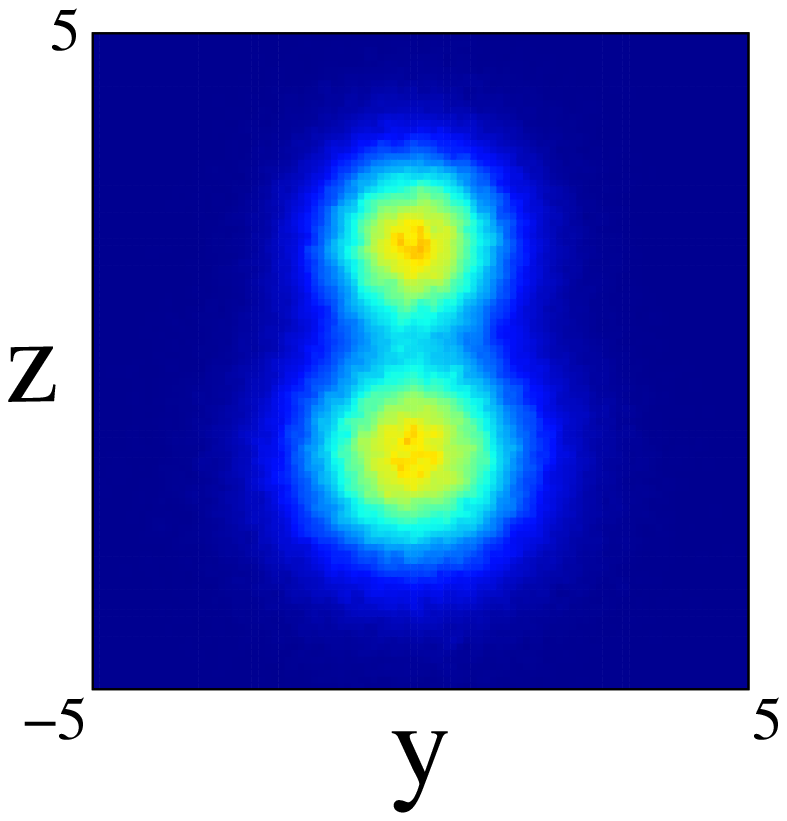}
\includegraphics[scale=0.28]{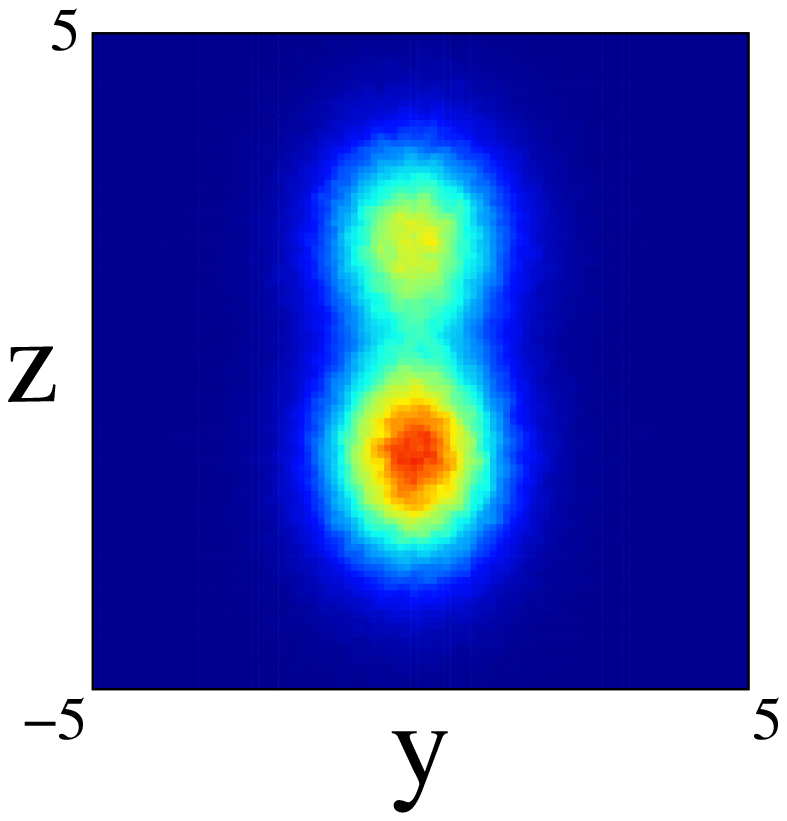}
\includegraphics[scale=0.28]{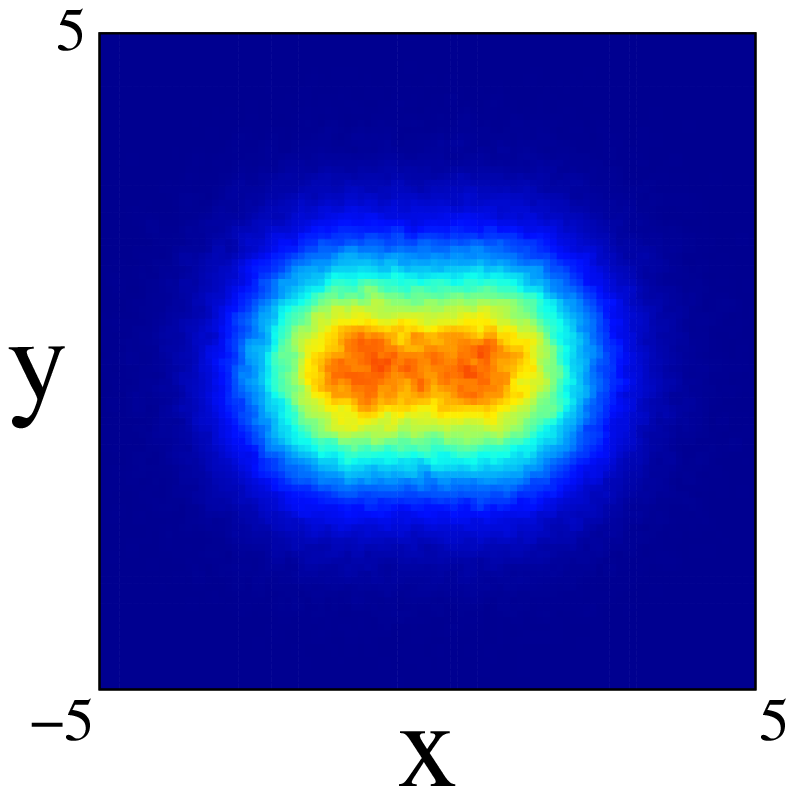}
\includegraphics[scale=0.28]{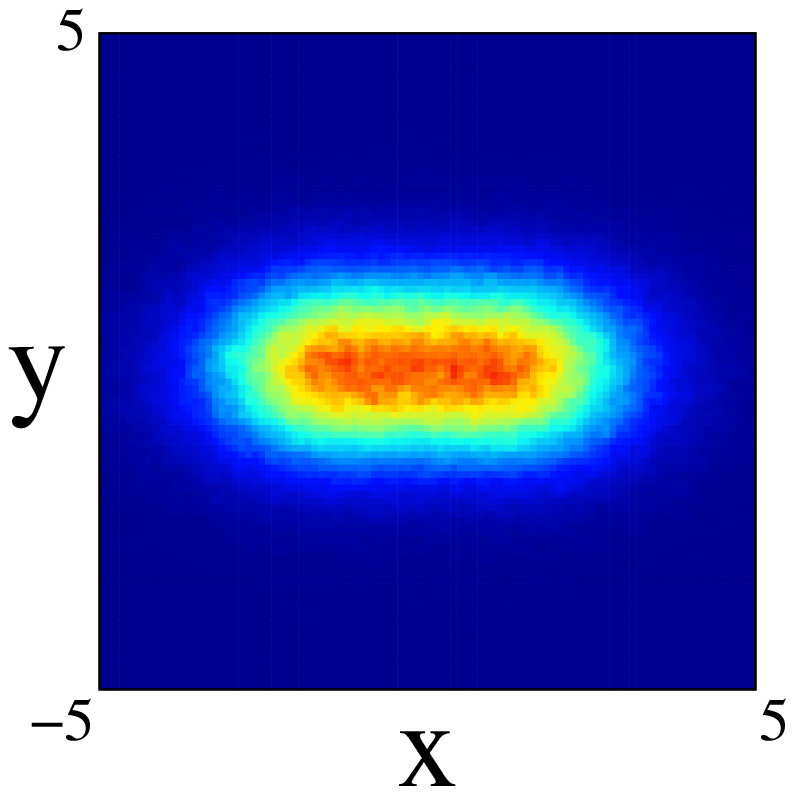}
\includegraphics[scale=0.28]{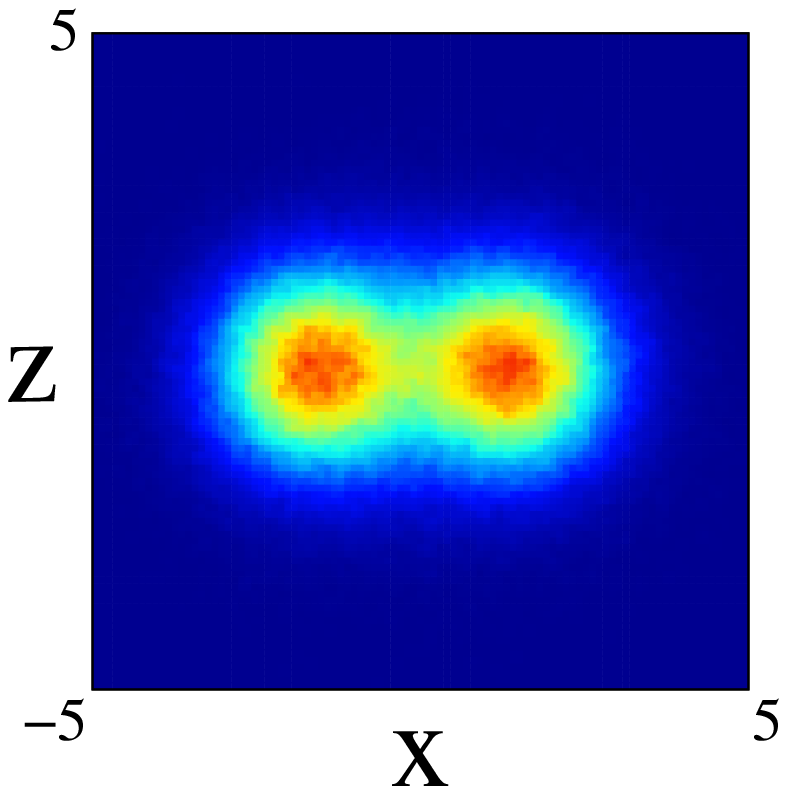}
\includegraphics[scale=0.28]{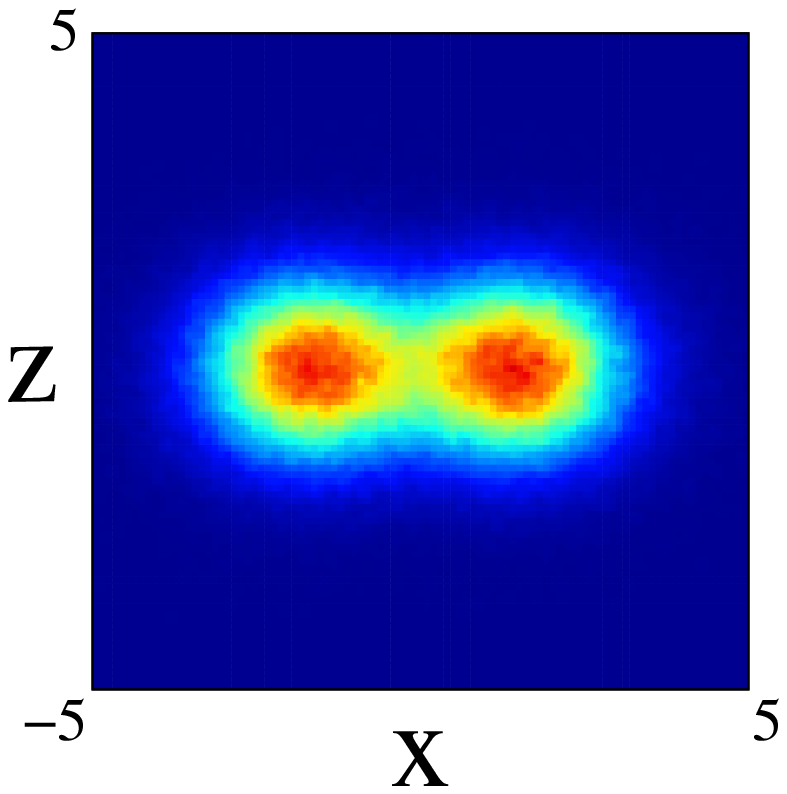}
\includegraphics[scale=0.28]{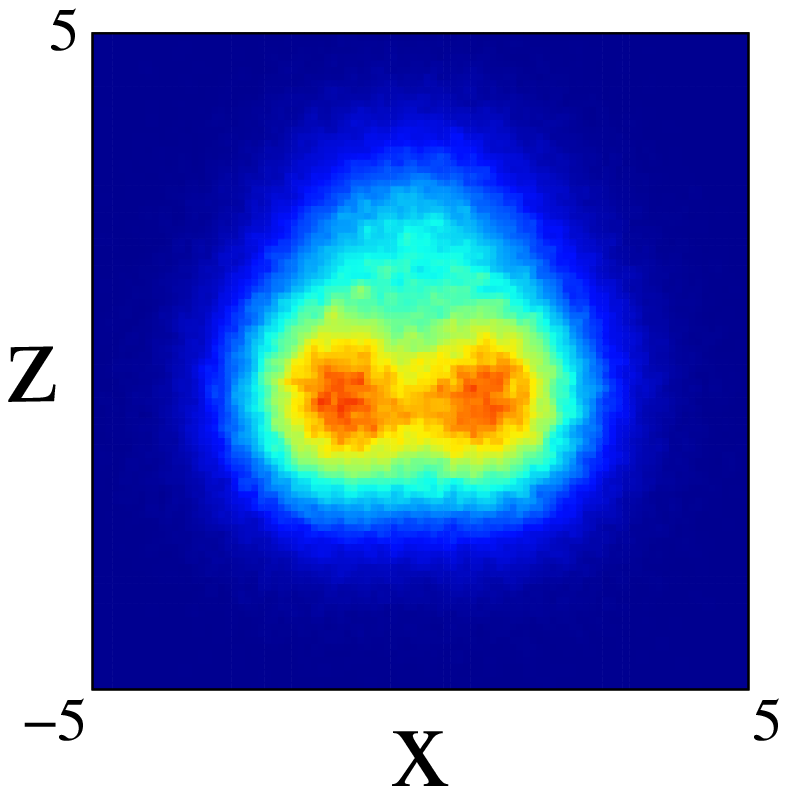}
\includegraphics[scale=0.28]{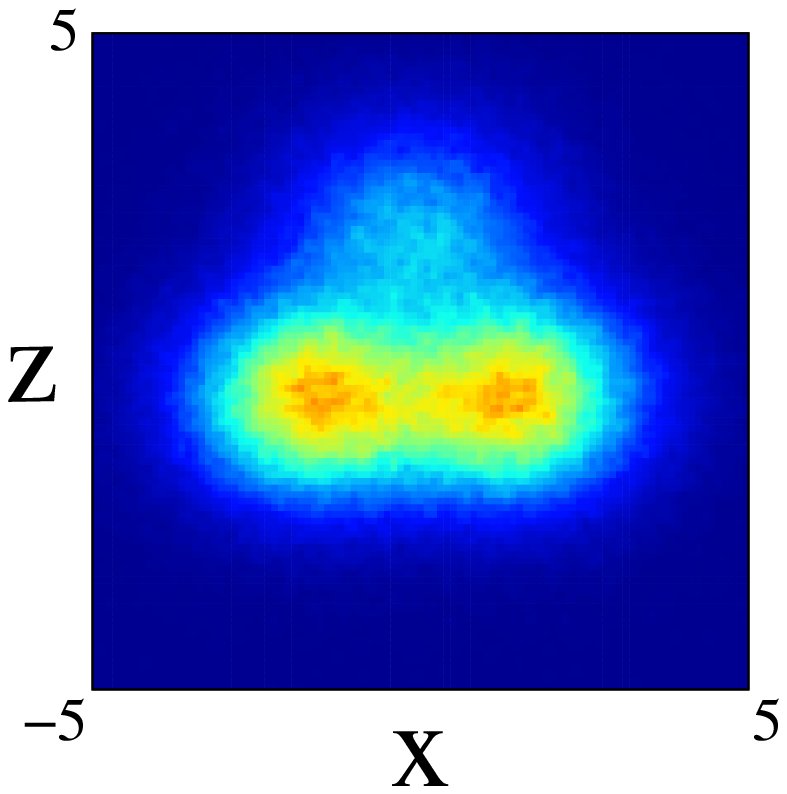}
\includegraphics[scale=0.28]{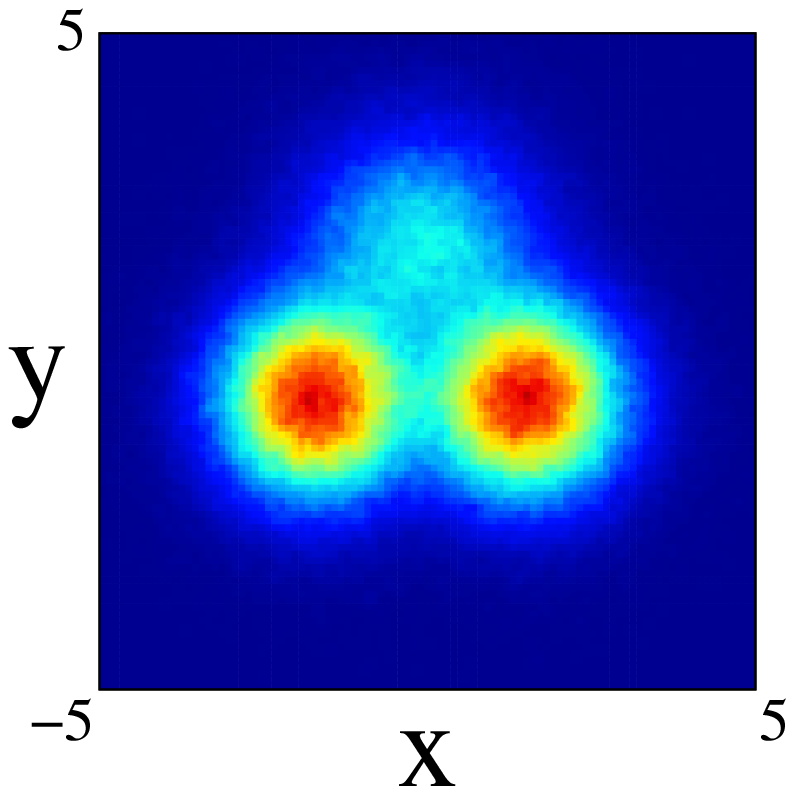}
\includegraphics[scale=0.28]{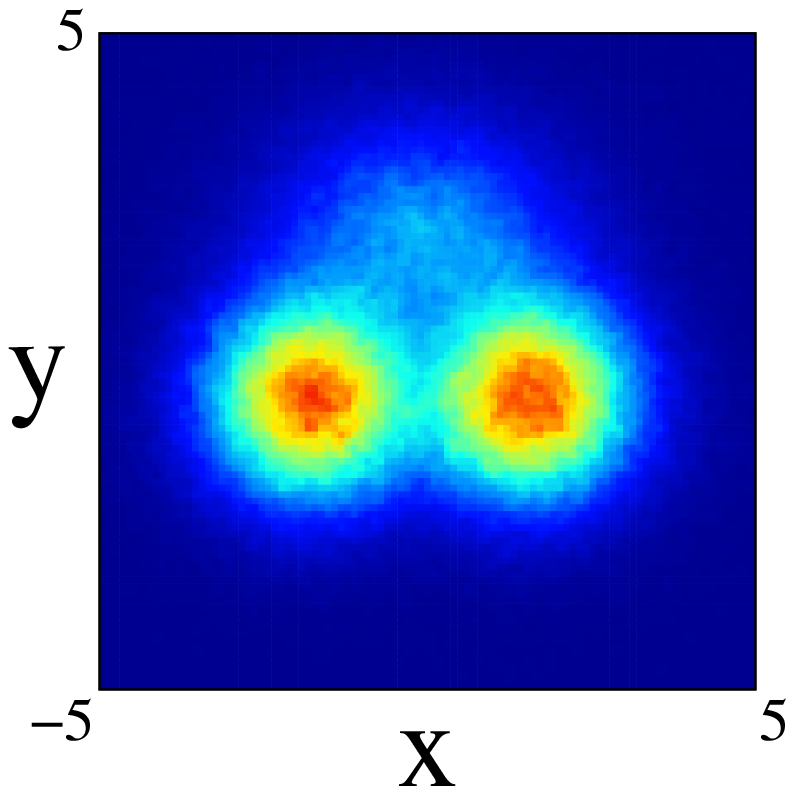}
\includegraphics[scale=0.28]{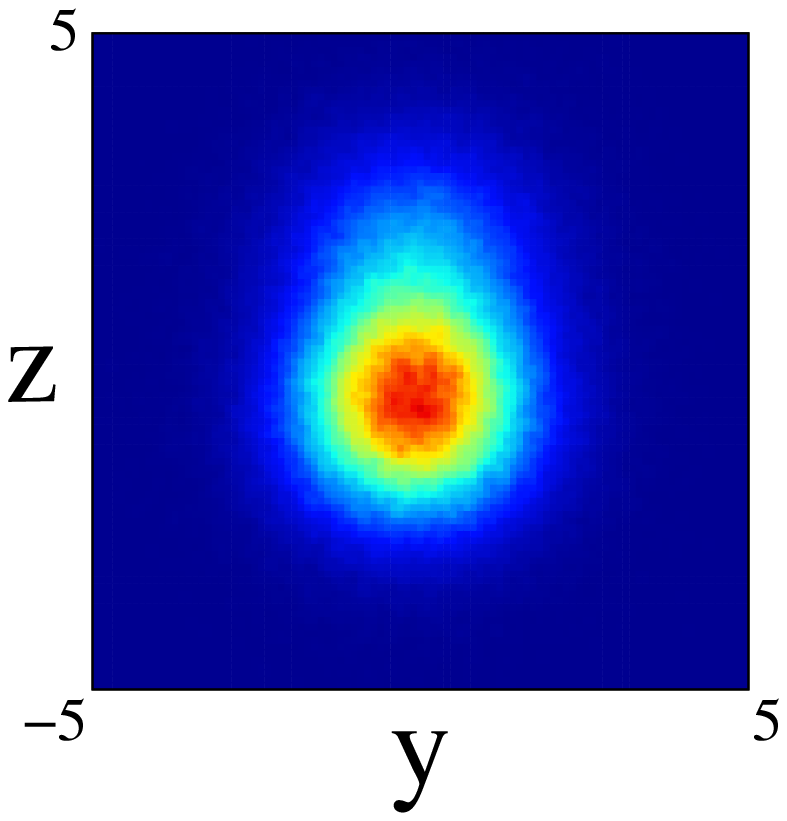}
\includegraphics[scale=0.28]{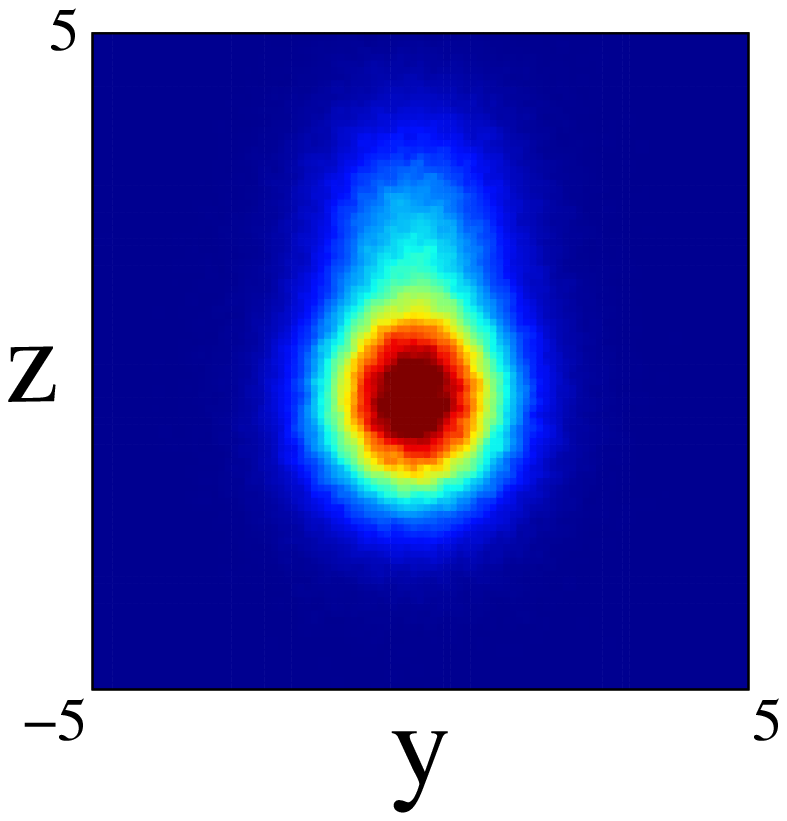}
\includegraphics[scale=0.28]{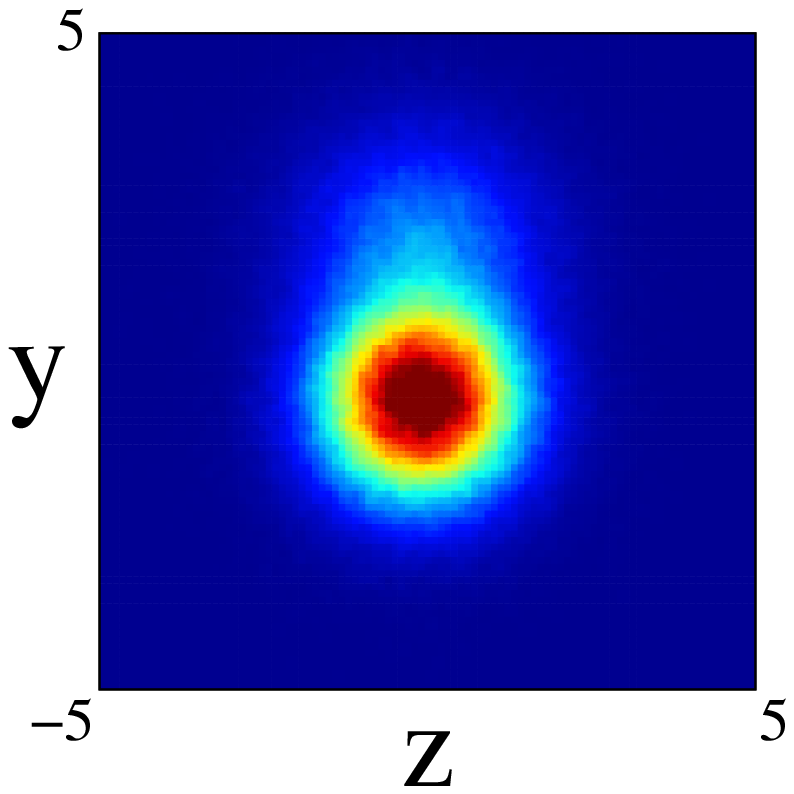}
\includegraphics[scale=0.28]{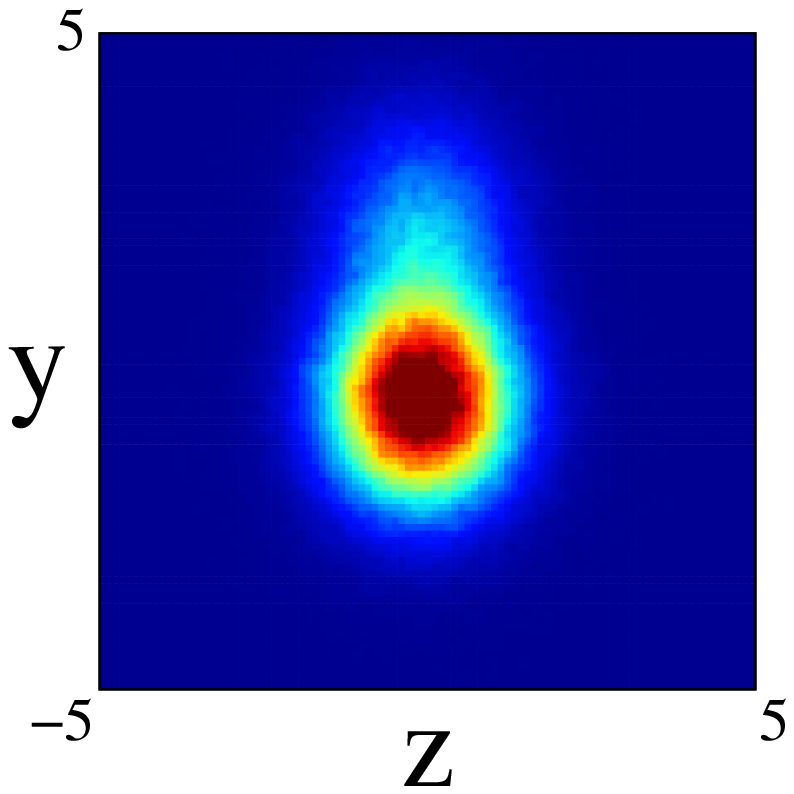}
\caption{\label{fig5} {\scriptsize $^{10}Be$ nucleus: Projection on the three
    cartesian planes formed by the axes shown as labels in the graphs 
    obtained for the distributions: (1) first two
    columns and first three rows; (3) last two columns and first three
    rows; (4) first two columns and last three rows and  (2)
    last two columns and last three rows.  For every of the
    distributions, the left column is within spherical approximation
    and the right one  is within the non-spherical. The usual planes
    disposition by rows is $xy$, $xz$ and $yz$ except for (2) where for
    the sake of comparison, $y$ and $z$ have been exchanged.}}
\end{center}
\end{figure}

Finally we show the results for the $^{10}$Be nucleus. In Figure
\ref{fig5}, we show the results of the projection for the four distributions in both
approximations and on the three cartesian planes (see figure caption
for further details). The locations of the two neutrons is quite clear
for all the distributions except the one center case, (1), where
appers as a diffuse halo in the $xy$ and $xz$ projections. It is
interesting that clusterization is clear even in some of the
distributions with an important part of the nucleons moving around the same
center. The spatial locations of the nucleons are different for every
of the distributions although there is an important resemblance
between (2) and (4). Comparing for every distribution, the results
between spherical and non-spherical approximations, there is a greater
diffusivity in the non-spherical approximation in the different
planes. However these differences almost disappear in the radial
density after promediating in the angles. 

\section{Conclusions.}

In the $^{10}$Be nucleus, the four spatial distributions studied are
adequate candidates for describing the ground state. This suggests
that a mixture of configurations with the four distributions could
provide a better approximation to the experimental results.

\section*{Appendix: Multidimensional gaussian integral.}

The integrals that appear in the calculation of matrix elements of the
operators considered in this work, and that may involve one or more
particles, can be written in general using cartesian coordinates as
\begin{equation}
\label{a1.1}
J(M,\vec{b};m_1,...,m_n) = 
\int_{\mathbb{R}^n}\prod_{k=1}^n dx_k~x_k^{m_k}
e^{-\vec{r} \cdot M \vec{r}+\vec{b}\cdot\vec{r}} \, \, ,
\end{equation}
with $\vec{b}$ is a fixed vector in the $n$--dimension space and $M$
is a non--singular $n \times n$ matrix, usually non symmetric.

Everyone of these integrals can be determined by direct derivation
respect to the components, $b_i$ of $\vec{b}$ from the integral:
\begin{equation}
\label{a1.2}
J\left(M,\vec{b};0,\ldots,0\right) =  
\int_{\mathbb{R}^n}
\prod_{k=1}^n dx_k~e^{-\vec{r}\cdot M\vec{r}+\vec{b}\cdot\vec{r}}
=\sqrt{\frac{\pi^{n}}{|M|}}~
e^{\frac 1 4 \vec{b}\cdot {M}^{-1}\vec{b}} \, \,  ,
\end{equation}
with $M^{-1}$ is the inverse matrix of $M$ and $|M|$
 its determinant. Derivating under the integral, it is easy to get
 that
\begin{equation}
\label{a1.4}
J(M,\vec{b};m_1,...,m_n) =  
\left\{\prod_{k=1}^n \frac{\partial^{m_k}}{\partial b_k^{m_k}}\right\}
J(M,\vec{b};0,\ldots,0) \, \, .
\end{equation}
Up to calculate all the integrals, we plan to establish a recurrence
relation among them. For that, it is useful to define:
\begin{eqnarray}
\label{a1.5a}
Q_j & = &  \sum_{i=1}^n b_iQ_{ji}\, \, ,\\
Q_{ij} & = & \frac 1 4 (M^{-1}_{ij}+M^{-1}_{ji})\, \, .\label{a1.5b}
\end{eqnarray}
Let us note that $Q_{ji}=Q_{ij}$ and independent of $\vec{b}$ while 
$Q_j$ depends linearly on $\vec{b}$ so  $\partial
Q_k/\partial b_l = Q_{kl}$.  Hereafter we shall write $J(m_1,...,m_n)$
instead of $J(M,\vec{b};m_1,...,m_n)$. It is easy to get in the case
when only the index $i$ is one and the rest are zeros that:
\begin{equation}
J(\ldots,1,\ldots) =\frac{\partial}{\partial b_i} J(0,\ldots,0)= Q_i
J(0,\ldots,0) \, \, .
\end{equation}
Taking this into account, we can write that:
\begin{equation}
J(m_1,\ldots,m_i +1,\ldots,m_n) =  
\left\{\prod_{k=1}^n \frac{\partial^{m_k}}{\partial b_k^{m_k}}\right\}
\left( Q_i J(0,\ldots,0) \right) \, \, ,
\end{equation}
and due to the mentioned properties of $Q_i$ and $Q_{ij}$, we can
finally get that:
\begin{eqnarray}
J(m_1,\ldots,m_i +1,\ldots,m_n) & = & Q_i
 J(m_1,\ldots,m_i,\ldots,m_n)+ \nonumber \\
& & \sum_{j=1}^n m_j  Q_{ij} J(m_1,\ldots,m_j-1,\ldots,m_n) \, \, .
\end{eqnarray}
This is the general recurrence relation that obliges to the index $i$
to be different from zero. A possible way of building
all the integral is beginning with $J(0,\ldots,0,l_n)$ from $l_n=1$ to
$l_n=m_n$ taking $i=n$. After that $J(0,\ldots,0,l_{n-1},l_n)$ taking
$i=n-1$ first $l_{n-1}=1$ and increase $l_n$ from zero to $m_n$ and 
so forth to get $l_{n-1}=m_{n-1}$ for all the
values of $l_n$. We continue adding a new index different from zero  that we take as
$i$ value until we finally get to 
$J(l_1,\ldots,l_n)$ with $i=1$.

\section*{Acknowledgments.} 

This work has been partially supported by the Spanish
Direcci\'on General de Inves\-tigaci\'on Cient\'{\i}fica y
T\'ecnica (DGICYT) under contract PB98--1318 and by the
Junta de Andaluc\'{\i}a.


\begin{thebibliography}{99}
\bibitem{Whee-1937} J.A. Wheeler, Phys. Rev. {\bf 52}, 1083 (1937).
\bibitem{Marg-1941} H. Margenau, Phys. Rev. {\bf 59}, 37 (1941).
\bibitem{GrWh-1957} J.J. Griffin and J.A. Wheeler, Phys. Rev. {\bf
    108}, 311 (1957).
\bibitem{Brin-1967} D.M. Brink, in: Proc. Int. School of Physics,
  Enrico Fermi 36, Academic Press, New York, p. 247 (1966).
\bibitem{BrWe-1968} D.M. Brink and A Weiguny: Nucl. Phys. {\bf
    A120}, 59 (1968).
\bibitem{AHT-1973} Y. Abe, J. Hiura and H. Tanaka,
  Prog. Theor. Phys. {\bf 49}, 800 (1973).
\bibitem{KHO-1995} Y. Kanada-En'yo, H. Horiuchi, and A. Ono,
  Phys. Rev. {\bf C52}, 628 (1995).
\bibitem{DeDu-1997} P. Descouvemont and M. Dufour, Nucl. Phys. {\bf A621}, C311 (1997).
\bibitem{DuDe-1997} M. Dufour and P. Descouvemont, Phys. Rev. {\bf C56},
  1831 (1997); Nucl. Phys. {\bf A726}, 53 (2003).
\bibitem{KaHo-1995} Y. Kanada-En'yo and H. Horiuchi, Phys. Rev. 
{\bf C52}, 647 (1995).
\bibitem{KHD-1999} Y. Kanada-En'yo, H. Horiuchi and A Dot\'e,
  Phys. Rev. {\bf C60}, 064304 (1999).
\bibitem{KaHo-2002} Y. Kanada-En'yo and H. Horiuchi,
  Phys. Rev. {\bf C66}, 024305 (2002).
\bibitem{LKT-1981} Q.K.K. Liu, H. Kanada and Y.C. Tang, 
  Phys. Rev. {\bf C23}, 645 (1981).
\bibitem{Desc-2004} P. Descouvemont, Phys. Rev. {\bf C70},
  065802 (2004).
\bibitem{DuDe-2008} M. Dufour and P. Descouvemont, Phys. Rev. {\bf C78},
  015808 (2008).
\bibitem{DeDu-2012} P. Descouvemont and M. Dufour, 'Microscopic
  Cluster Models' pags 1-66, in 'Cluster in Nuclei, Vol.2' edited by 
  C. Beck, Lectures Notes in Physics {\bf 848}, Springer-Verlag (2012).
\bibitem{OFK-2006} W. von Oertzen, M. Freer, Y. Kanada-En’yo,
  Phys. Rep. {\bf 432}, 43 (2006)
\bibitem{DuDe-1996} M. Dufour and P. Descouvemont,
  Nucl. Phys. {\bf A605}, 160 (1996).
\bibitem{EiGr-1988} J.M. Eisenberg and W. Greiner, Nuclear Theory,
  Volume 1: Nuclear Models, Third edition, North Holland (1988)
\bibitem{Volk-1965} A. B. Volkov, Nucl. Phys. {\bf 74}, 33 (1965).
\bibitem{BrBo-1967} D. Brink and E. Boeker, Nucl. Phys. {\bf 91}, 1 (1967).
\bibitem{TLT-1977} D. R. Thomson, M. Lemere and Y. C. Tang,
 Nucl. Phys. {\bf A286}, 53 (1977).
\bibitem{Desc-2002} P. Descouvemont,  Nucl. Phys. A699, 463 (2002). 
\end{thebibliography}
\end{document}